\documentclass[11pt]{article}

\usepackage{amsmath,amssymb,amsthm,mathtools}
\usepackage[margin=1in]{geometry}
\usepackage{booktabs}
\usepackage{graphicx}
\usepackage{tikz}
\usepackage{microtype}
\usepackage{mathrsfs}
\usepackage{enumitem}
\usepackage{hyperref}
\usepackage{setspace}
\hypersetup{colorlinks=true,linkcolor=blue,citecolor=blue,urlcolor=blue}
\newtheorem{theorem}{Theorem}[section]
\newtheorem{lemma}[theorem]{Lemma}
\newtheorem{proposition}[theorem]{Proposition}
\newtheorem{corollary}[theorem]{Corollary}
\newtheorem{remark}[theorem]{Remark}
\theoremstyle{definition}
\newtheorem{definition}[theorem]{Definition}
\newtheorem{assumption}[theorem]{Assumption}

\newcommand{\R}{\mathbb{R}}
\newcommand{\E}{\mathbb{E}}
\newcommand{\Pbb}{\mathbb{P}}

\newcommand{\mmin}{\underline{m}}
\newcommand{\ind}[1]{\mathbf{1}_{#1}}

\title{Robust bounds for American fixed-strike lookback puts }
\author{YongQing Xu \\ \rm  \small  School of Artificial Intelligence,  Shenzhen Technology University, Shenzhen~518118, China
}
\date{ }

\begin{document}
\maketitle

\begin{abstract}
We initiate the robust (model-independent) pricing of American lookback options
under martingale optimal transport (MOT) constraints. Given finitely many
marginals of the discounted price, we study the maximal expected discounted
payoff of an American \emph{fixed-strike lookback put}, whose exercise value at
$t$ is $(K-\min_{u\le t}e^{ru}X_u)^+$, i.e.\ a function of the running minimum
$m_t$ of the undiscounted price. Two structural facts isolate this instrument as
the non-trivial American lookback problem. First, at zero interest rate every
classical lookback degenerates to its European counterpart (the discounted payoff
is a submartingale). Second, at $r>0$ the \emph{max}-based fixed-strike lookback
call remains European-type (the running maximum is unbounded and is attained
late), whereas the \emph{min}-based fixed-strike lookback put is genuinely
American (the running minimum is bounded below and is attained early). We prove a
pricing--hedging duality, applying the continuous-time duality of
Guo--Ob\l\'oj together with a reduction of the hedge to the two-dimensional state
$(X_t,m_t)$. In the two-period model we establish a convexity lemma and an exact
decomposition of the continuation payoff into a locked intrinsic part plus a
\emph{minimum-drop put}, from which we derive the exercise criterion and the
closed-form free-continuation value. The optimal continuation coupling is then
identified as the \emph{monotone martingale coupling}---the third extremal
martingale coupling, with both shadow functions non-decreasing, distinct from the
left- and right-curtain couplings---and we prove, and confirm numerically by
linear programming, that the optimal exercise region is a \emph{right tail}
$\{x\ge x^*\}$, in sharp contrast to the left-tail exercise region of the
standard American put. The threshold $x^*$ is monotone in the parameters
(increasing in volatility, decreasing in $r$, and diverging as $r\to0$ with the
American premium vanishing), and is affine in $K$ with a $K$-independent shadow
pair for $K\ge s_0$. The structure extends to finitely many marginals
\emph{unconditionally}: at every stage the optimal exercise region is a right
tail, and the optimal coupling is the reflected iterated Az\'ema--Yor embedding
(the reflection of the maximum--maximum embedding of
Henry-Labord\`ere--Ob\l\'oj--Spoida--Touzi), which minimises the running minimum in
stochastic order. Under the reflected Ob\l\'oj--Spoida Assumption $\circledast$
this embedding is the two-point iterated monotone martingale coupling; when
$\circledast$ fails it carries the mass-diffusion correction of
Ob\l\'oj--Spoida, which we show numerically to be \emph{local} (consistent under
projection), reducing the $n>3$ problem to the iteration of the explicit $n=3$
correction. We quantify the American and model-risk premia.
\end{abstract}



\section{Introduction}\label{sec:intro}

Martingale optimal transport (MOT) prices and hedges a derivative robustly
against the class of all martingale models consistent with finitely many
observed option prices. For European claims this is by now a mature theory: the
underlying duality goes back to \cite{Strassen65,BJ2012,BNT2015}, the extremal
(``curtain'' and monotone) couplings to \cite{BJ2016,BHLT2017}, and the lookback
and longest-running-maximum bounds to \cite{Hobson98,HOST2016}, with
\cite{GHT14,HOb2016} covering related path-dependent robust bounds. The robust
pricing of \emph{American} claims is more delicate: the American price is a
supremum over both models \emph{and} stopping times, which do not commute.
Hobson--Neuberger \cite{HN2017} proved that the model-free supremum is attained
by a regime model, and Hobson--Norgilas \cite{HobNor2019} obtained explicit
two-period bounds for the American put, whose optimal exercise region is a
\emph{left} tail---the familiar early-exercise-at-low-price phenomenon.

This paper takes the intersection of these two lines---American robust bounds
$\times$ lookback path functionals---which appears not to have been treated
before. We focus on the instrument that is the genuinely American one in this
class, the \emph{fixed-strike lookback put}.

\begin{definition}[Fixed-strike lookback put]\label{def:payoff}
For a discounted price $X$ with $X_0=s_0$ and interest rate $r>0$, the
\emph{fixed-strike lookback put} pays
\[
  Z(t,\omega)=e^{-rt}\bigl(K-\mmin_t(\omega)\bigr)^+,
  \qquad
  \mmin_t(\omega)=\min_{0\le u\le t} e^{ru}X_u(\omega),
\]
when exercised at time $t$.
\end{definition}
\begin{figure}[htbp]
\centering
\includegraphics[width=0.65\textwidth]{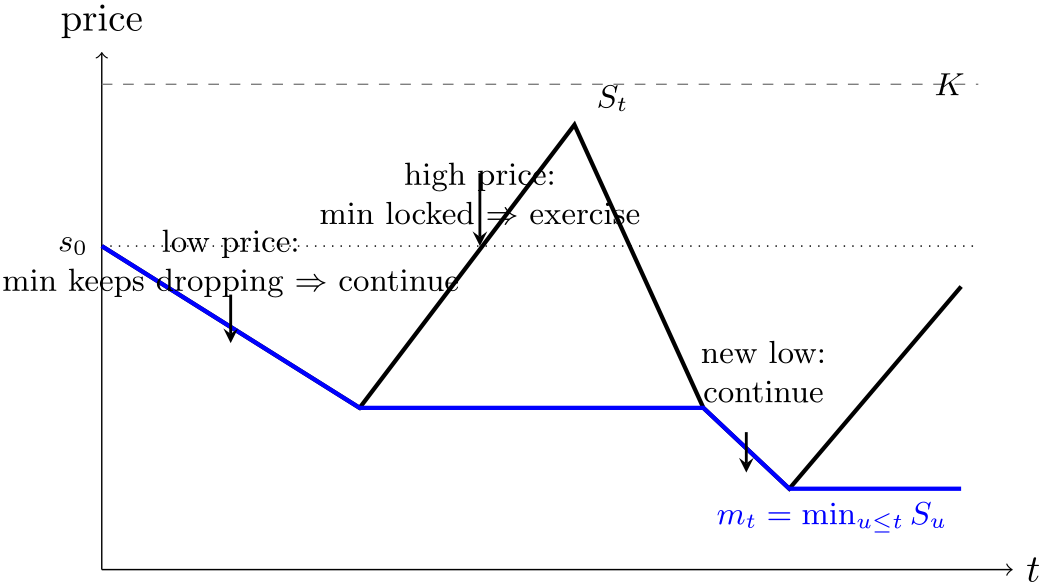}
\caption{Why the exercise region is a right tail. A sample price path $S_t$ and
its running minimum $m_t=\min_{u\le t}S_u$. The minimum only ratchets
\emph{down}: low prices keep the minimum dropping (continuation is valuable),
while high prices leave the minimum locked, so waiting only loses the discount
factor and immediate exercise locks in the intrinsic value $(K-m_t)^+$.}
\label{fig:ratchet}
\end{figure}
We isolate it by a \emph{structural degeneracy}. At $r=0$ all four classical
lookbacks collapse to their European values (the discounted payoff is a
submartingale, Proposition~\ref{prop:r0}), so there is nothing American to price.
At $r>0$ the \emph{max}-based fixed-strike lookback call remains European-type
(the running maximum is unbounded and is attained late), whereas the
\emph{min}-based put is genuinely American (the running minimum is bounded below
and is attained early, Proposition~\ref{prop:asym}). Throughout the rest of the
paper $r>0$.

\paragraph{Our contributions.}
\begin{enumerate}[label=(\arabic*),leftmargin=2em]
\item \emph{Pricing--hedging duality.} Theorem~\ref{thm:A} applies the
continuous-time duality of \cite{GuoObloj2025} to the running-minimum payoff and
reduces the hedge to the two-dimensional state $(X_t,m_t)$
(Lemma~\ref{lem:markov}).

\item \emph{Two-period structure.} A convexity lemma
(Lemma~\ref{lem:convex}) and an exact decomposition of the continuation payoff
into a locked intrinsic part plus a \emph{minimum-drop put}
(Lemma~\ref{lem:decomp}) yield the exercise criterion
(Proposition~\ref{prop:criterion}) and the closed-form free continuation value
(Proposition~\ref{prop:cfree}), and reduce the problem to a scalar
martingale-optimal-transport problem (Lemma~\ref{lem:reduction}).

\item \emph{Right-tail exercise and the third extremal coupling.} The main
structure theorem (Theorem~\ref{thm:B}) establishes that the optimal exercise
region is a \emph{right} tail $\{x\ge x^*\}$---the \emph{opposite} of the
left-tail exercise region of the standard American put, because the running
minimum ratchets only \emph{down}. The optimal continuation coupling is the
\emph{monotone martingale coupling} of \cite{BHLT2017}
(Proposition~\ref{prop:lc}), the third extremal martingale coupling---distinct
from the left- and right-curtain couplings, with \emph{both} shadows
non-decreasing---identified here by an envelope-theorem argument rather than by
submodularity (Remark~\ref{rem:terminology}).

\item \emph{Comparative statics.} The threshold $x^*$ is monotone in the
parameters and affine in $K$ with a $K$-independent shadow pair for $K\ge s_0$
(Proposition~\ref{prop:compstat}): $x^*$ increases in volatility, decreases in
$r$, and diverges as $r\to0$ while the American premium vanishes.

\item \emph{Multi-marginal extension and mass diffusion.} Under the reflected
Ob\l\'oj--Spoida Assumption~\ref{ass:star} the structure extends to finitely
many marginals: the optimal coupling is the reflected iterated Az\'ema--Yor
embedding and the exercise region is a right tail at every stage
(Theorem~\ref{thm:multi}). Without the assumption the coupling carries the
\emph{mass-diffusion} correction of \cite[\S4.2]{ObSpoida2015}, which we detect
numerically for $n=3$ (Remark~\ref{rem:n3num}) and $n=4$
(Remark~\ref{rem:telescope}); the correction is \emph{local} (consistent under
projection, Lemma~\ref{lem:consistent}), so the explicit $n>3$ problem reduces to
iterating the explicit $n=3$ correction.
\end{enumerate}

\paragraph{Organisation.} Section~\ref{sec:setup} establishes the structural
degeneracy; Section~\ref{sec:two} sets up the two-period entry problem;
Section~\ref{sec:duality} proves the duality and the Markovian reduction;
Section~\ref{sec:structure} develops the two-period optimal structure and the
comparative statics; Section~\ref{sec:numerics} contains the LP experiments; and
Section~\ref{sec:multi} carries the analysis to $n$ marginals.

\subsection{Notation}
$\mu\le_{\mathrm{cx}}\nu$ denotes convex order. $\Pi_m(\mu,\nu)$ is the set of
martingale couplings of $\mu$ and $\nu$. Throughout, $d_1=e^{-rt_1}$,
$d_2=e^{-rt_2}$, and $m_1(x)=\min(s_0,e^{rt_1}x)$.

\section{Setup and the structural degeneracy}\label{sec:setup}

Let $\Omega=C([0,T];\R_+)$ be the canonical space of continuous discounted price
paths $X=(X_t)$, with $X_0=s_0$. Pricing measures are probability measures under
which $X$ is a martingale. The undiscounted price is $S_t=e^{rt}X_t$, a process
with upward drift for $r>0$.

\subsection{Zero interest rate}

\begin{proposition}\label{prop:r0}
Let $r=0$. For each of the four classical lookbacks
\[
  (M_t-K)^+,\quad M_t-X_t,\quad (K-m_t)^+,\quad X_t-m_t,
\]
where $M_t=\max_{u\le t}X_u$, $m_t=\min_{u\le t}X_u$, the discounted payoff
process is a submartingale, so that early exercise is never strictly optimal:
the robust American price equals the robust European price.
\end{proposition}

\begin{proof}
At $r=0$, $X$ is a martingale, $M$ is nondecreasing and $m$ is nonincreasing.
For $(M_t-K)^+$: $x\mapsto(x-K)^+$ is convex nondecreasing, hence
$(M_t-K)^+$ is a submartingale. For $M_t-X_t$: $M$ is a submartingale and $X$ a
martingale, so the difference is a submartingale. For $(K-m_t)^+$:
$x\mapsto(K-x)^+$ is convex nonincreasing and $m$ is a supermartingale, hence
$(K-m_t)^+$ is a submartingale. For $X_t-m_t$: $-m$ is a submartingale, so
$X-m$ is one. Optional sampling gives $Z_\tau\le \E[Z_T\mid\mathcal F_\tau]$ for
every stopping time $\tau$.
\end{proof}

\subsection{Positive interest rate: the call/put asymmetry}

\begin{proposition}\label{prop:asym}
At $r>0$ the fixed-strike lookback \emph{call} $(M_t-K)^+$ is European-type: its
robust American price equals its robust European price for every $K,r$. In
contrast the fixed-strike lookback \emph{put} $(K-m_t)^+$ is genuinely American:
its American premium is strictly positive for $K\ge s_0$ and $r>0$, and is
monotone increasing in $r$ and $K$.
\end{proposition}

The mechanism is the upward drift of $S_t=e^{rt}X_t$: the running maximum is
attained \emph{late} (so the call waits), while the running minimum is attained
\emph{early} (so the put may exercise to lock in a low minimum before it is
eroded by discounting). The LP experiments of Section~\ref{sec:numerics} confirm
this: the call American premium is identically zero over all $(K,r)$ tested,
while the put premium grows in $r$ and $K$ (Table~\ref{tab:rsweep}).

Consequently the genuine gap is the \emph{American fixed-strike lookback put},
with two-dimensional state $(X_t,m_t)$; we treat this problem from here on.

\section{The two-period entry problem}\label{sec:two}

Fix times $t_1<t_2$, marginals $\mu=\mathrm{Law}(X_{t_1})$,
$\nu=\mathrm{Law}(X_{t_2})$ in convex order with common mean $s_0$. The
\emph{robust} (model-free) price of the American lookback put is
\begin{equation}\label{eq:primal}
  P \;=\; \sup_{\pi\in\Pi_m(\mu,\nu)}\;\sup_{B\subseteq\R_+}
  \E_\pi\!\Big[\, a_1(X)\,\ind{X\in B} \;+\; a_2(X,Y)\,\ind{X\notin B}\Big],
\end{equation}
where
\begin{equation}\label{eq:payoffs}
  a_1(x)=d_1\bigl(K-m_1(x)\bigr)^+,\qquad
  a_2(x,y)=d_2\bigl(K-\min(m_1(x),e^{rt_2}y)\bigr)^+,
\end{equation}
and $m_1(x)=\min(s_0,e^{rt_1}x)$ is the running minimum at time $t_1$. Note the
structural difference from the standard American put of \cite{HN2017,HobNor2019}:
the continuation payoff $a_2(x,y)$ is genuinely bivariate (it depends on $x$
through the running minimum $m_1(x)$), whereas the standard put continuation
payoff $(K_2-y)^+$ depends on $y$ alone. This is the entire difficulty, and the
source of the novel structure.

\section{Duality and the Markovian reduction}\label{sec:duality}

The following is the running-min specialisation of the continuous-time
pricing--hedging duality of Guo--Ob\l\'oj \cite{GuoObloj2025}.

\begin{theorem}[Robust pricing--hedging duality]\label{thm:A}
Let $Z(t,\omega)=e^{-rt}(K-m_t(\omega))^+$ and let $\mathcal Q^g_E$ be the set of
calibrated martingale measures (volatility restriction $E$, static instruments
$g$). Then
\[
  \sup_{P\in\mathcal Q^g_E,\ \tau\in\mathcal T^P}\E^P[Z_\tau]
  \;=\;\pi^A_{g,E}(Z),
\]
where $\pi^A_{g,E}(Z)$ is the robust superhedging price
\[
  \inf\Big\{\, x : \exists\,(q,\tilde q,h)
  \ \ x+\int_0^u q_s\,\mathrm dX_s+\int_u^1 \tilde q_u(s)\,\mathrm dX_s
  +h\!\cdot\! g \ge Z(u,\cdot)\ \ \forall u,\ \mathcal Q^g_E\text{-q.s.}\Big\}.
\]
Moreover, since $Z$ depends on the path only through $(X,m)$, the optimal hedge
can be taken of Markov form $q_s=q(s,X_s,m_s)$, i.e.\ a function of the
two-dimensional state $(X_t,m_t)$.
\end{theorem}

\begin{remark}\label{rem:guo}
The payoff $Z(t,\omega)=e^{-rt}(K-m_t(\omega))^+$ satisfies exactly the standing
hypothesis of \cite[Thm.~5.2]{GuoObloj2025}: it is bounded ($0\le Z\le K$) and
$m_t$ is a continuous functional of the stopped path $\omega_{\cdot\wedge t}$, so
$Z\in C_b(\Lambda)$ and the no-gap duality of Theorem~\ref{thm:A} is an immediate
application. The genuinely new content of the theorem is its second sentence, the
Markov reduction of the hedge, which we now make precise and prove.
\end{remark}

\begin{lemma}[Markovian value and hedge]\label{lem:markov}
Under the standing hypotheses of Theorem~\ref{thm:A} the following hold.
\begin{enumerate}
\item[(i)] For every $P\in\mathcal Q^g_E$ the pair $(X_t,m_t)$ is a
  (time-inhomogeneous) Markov process, and the discounted value process
  \begin{equation}\label{eq:snell}
    V_t=\operatorname{ess\,sup}_{\tau\in\mathcal T_t^P}
    \E^P\!\bigl[e^{-r(\tau-t)}(K-m_\tau)^+\bigm|\mathcal F_t\bigr],
    \qquad \mathcal T_t^P=\{\tau\in\mathcal T^P:\tau\ge t\},
  \end{equation}
  is of the form $V_t=V(t,X_t,m_t)$ for a deterministic Borel function
  $V:[0,T]\times\R_+\times[0,s_0]\to\R_+$.
\item[(ii)] The dynamic strategy in the superhedging representation of
  Theorem~\ref{thm:A} can be taken Markovian, $q_s=q(s,X_s,m_s)$ (and likewise
  for the nested term $\tilde q$). On the region where $V\in C^{1,2}$ one has
  $q(s,x,m)=\partial_x V(s,x,m)$.
\end{enumerate}
\end{lemma}

\begin{proof}
(i) Write $S_t=e^{rt}X_t$, so that $m_t=\min_{u\le t}S_u$. Under $P$ the process
$S$ is a Markov process (a bijection of the martingale $X$), and the
running-minimum pair $(S,m)$ is the standard reflected pair
$m_{t+s}=\min(m_t,\inf_{u\in[t,t+s]}S_u)$, hence Markov. Therefore
$(X_t,m_t)=(e^{-rt}S_t,m_t)$ is a (time-inhomogeneous) Markov process. The
reward $Z_\tau=e^{-r\tau}(K-m_\tau)^+$ is a function of $(\tau,X_\tau,m_\tau)$;
by the Markov property and the Snell-envelope fixed point
$V_t=\max\bigl(Z_t,\E^P[V_{t+}\mid\mathcal F_t]\bigr)$ the value process depends
on $\mathcal F_t$ only through $(t,X_t,m_t)$, proving (i).

(ii) The process $V$ is the Snell envelope of $Z$, a $P$-supermartingale; by the
Doob--Meyer decomposition $V_t=V_0+M_t-A_t$ with $M$ a martingale and $A$ a
predictable increasing process. The dynamic hedge of Theorem~\ref{thm:A} is the
martingale part $M$, and by (i) this is the supermartingale decomposition of the
Markov functional $V(t,X_t,m_t)$; consequently the integrand $q_s$ in
$M_t=\int_0^t q_s\,\mathrm dX_s$ inherits the Markov structure,
$q_s=q(s,X_s,m_s)$. For the identification: the running minimum $m$ is
non-increasing (finite variation) with $\mathrm dm_s\neq0$ only on the diagonal
$\{S_s=m_s\}$; on the region where $V\in C^{1,2}$, It\^o's formula gives
\[
  \mathrm dV(t,X_t,m_t)=\partial_t V\,\mathrm dt+\partial_x V\,\mathrm dX_t
  +\partial_m V\,\mathrm dm_t+\tfrac12\,\partial_{xx}V\,\mathrm d\langle X\rangle_t .
\]
As $X$ is a $P$-martingale, the only martingale term on the right-hand side is
$\partial_x V\,\mathrm dX_t$; the remaining terms are of finite variation and are
absorbed into the compensator $A$. Hence $q(s,x,m)=\partial_x V(s,x,m)$ on the
smooth region, and in general $q$ is the $x$-component of the supermartingale
decomposition of $V$, characterised by the standard viscosity
(variational-inequality) formulation of the Snell envelope. This proves (ii).
\end{proof}

\section{The two-period optimal structure}\label{sec:structure}

Throughout this section we impose the non-degenerate regime
(Remark~\ref{rem:degenerate} below) in which the optimal exercise decision is
strict.

\subsection{Convexity and a clean decomposition}

\begin{lemma}\label{lem:convex}
For fixed $x$, the continuation payoff $y\mapsto a_2(x,y)$ is decreasing and
convex on $\R_+$.
\end{lemma}

\begin{proof}
$y\mapsto\min(m_1(x),e^{rt_2}y)$ is concave increasing (a minimum of an affine
function and a constant), so $y\mapsto K-\min(m_1(x),e^{rt_2}y)$ is convex
decreasing, and the positive part of a convex decreasing function is convex
decreasing.
\end{proof}

\begin{lemma}[Decomposition]\label{lem:decomp}
For $m_1(x)<K$,
\begin{equation}\label{eq:decomp}
  a_2(x,y)\;=\;d_2\bigl(K-m_1(x)\bigr)\;+\;d_2\bigl(m_1(x)-e^{rt_2}y\bigr)^+.
\end{equation}
\end{lemma}

\begin{proof}
For $m_1(x)<K$,
\[
  \bigl(K-\min(m_1,e^{rt_2}y)\bigr)^+
  =(K-m_1)+(m_1-e^{rt_2}y)^+ ,
\]
which is verified directly by splitting on $e^{rt_2}y\gtrless m_1$.
\end{proof}

The decomposition \eqref{eq:decomp} is the structural core of the problem: the
continuation payoff splits into a \emph{locked intrinsic} part $d_2(K-m_1(x))$
(deterministic given $x$) and a \emph{minimum-drop put} with strike
$m_1(x)e^{-rt_2}$. Both parts depend on $x$ only through the running minimum
$m_1(x)$: the payoff structure collapses to the single variable $m_1$, and $x$
enters the coupling only through the martingale constraint
$\E[Y\mid x]=x$.

\begin{proposition}[Exercise criterion]\label{prop:criterion}
Write $\Delta=t_2-t_1$. Under any continuation coupling,
\[
  \text{continue is (weakly) optimal at } x
  \iff
  \E\!\bigl[(m_1(x)-e^{rt_2}Y)^+\bigr]
  \;\ge\;(e^{r\Delta}-1)\bigl(K-m_1(x)\bigr).
\]
\end{proposition}

\begin{proof}
Continue iff $d_2(K-m_1)+d_2\,\E[(m_1-e^{rt_2}Y)^+]\ge d_1(K-m_1)$; divide by
$d_2$ and use $d_1/d_2=e^{r\Delta}$.
\end{proof}

\subsection{The free continuation value}

Before imposing the global constraint, we compute the continuation value a single
node $x$ can achieve if it may use $\nu$'s tail without competing with other
nodes. This exposes why exercise is a \emph{global} scarcity effect, not a local
trade-off.

\begin{proposition}[Free continuation value]\label{prop:cfree}
For a single node $x$ whose continuation law is supported on $[a,b]$, the
supremum over all martingale kernels $q$ with $\E_q[Y]=x$ is
\begin{equation}\label{eq:cfree}
  c_{\mathrm{free}}(x)=\alpha\, a_2(x,a)+(1-\alpha)\,a_2(x,b),\qquad
  \alpha=\frac{b-x}{b-a}.
\end{equation}
\end{proposition}

\begin{proof}
Since $y\mapsto a_2(x,y)$ is convex (Lemma~\ref{lem:convex}), the supremum over
$q$ with mean $x$ is attained at an extreme point of the set of such kernels,
namely the two-point martingale coupling $q=\alpha\delta_a+(1-\alpha)\delta_b$,
where $\alpha$ is uniquely determined by $\alpha a+(1-\alpha)b=x$.
\end{proof}

\begin{corollary}[No local incentive to exercise]\label{cor:local}
For every $x$ with $m_1(x)<K$ and a sufficiently rich left tail of $\nu$,
$c_{\mathrm{free}}(x)>a_1(x)$; exercise is therefore \emph{never} locally
profitable. Exercise arises purely because the left tail of $\nu$ is scarce and
is allocated, by the global coupling, to the deep in-the-money low-$x$ nodes,
crowding the high-$x$ nodes out.
\end{corollary}

\begin{remark}[Numerical verification]\label{rem:cfree_num}
For the CRR marginals of Section~\ref{sec:numerics} ($s_0=100$, $r=0.05$,
$t_1=0.5$, $t_2=1$, $K=120$), at $x=61$ we have $m_1(x)=62.69$, $a\approx30$,
$b\approx354$, so $a_2(x,a)=84.1$, $a_2(x,b)=54.4$, $\alpha\approx0.904$, giving
$c_{\mathrm{free}}\approx81.2$, against $a_1(x)=55.9$. The free continuation
value dominates exercise by a factor $1.45$.
\end{remark}

\subsection{Reduction to a scalar martingale-optimal-transport problem}

Using the decomposition \eqref{eq:decomp}, the two-period problem collapses: the
continuation value of the whole exercise region is a fixed term plus a single
\emph{scalar} MOT problem. This reduction is the clean object that the
monotone martingale coupling of Proposition~\ref{prop:lc} must optimise.

\begin{lemma}[Reduction]\label{lem:reduction}
Let $K\ge s_0$ and fix an exercise threshold $x^*$. Then
\[
  P(x^*) \;:=\; \sup_{\pi\in\Pi_m(\mu,\nu)}
  \E_\pi\Big[a_1(X)\ind{X\ge x^*}+a_2(X,Y)\ind{X<x^*}\Big]
\]
satisfies
\[
  P(x^*) \;=\;
  \E_\mu\bigl[a_1(X)\ind{X\ge x^*}\bigr]
  \;+\; d_2\int_{(-\infty,x^*)}\bigl(K-m_1(x)\bigr)\,\mu(\mathrm dx)
  \;+\; d_2\,V(x^*),
\]
where
\begin{equation}\label{eq:scalar}
  V(x^*) \;=\; \sup_{\pi\in\Pi_m(\mu,\nu)}
  \E_\pi\Big[\bigl(m_1(X)-e^{rt_2}Y\bigr)^+\ind{X<x^*}\Big].
\end{equation}
On the continuation region the running minimum is a non-decreasing function of
$X$, so \eqref{eq:scalar} is a scalar MOT problem: writing
\[
  \alpha(x)=e^{-r\Delta}x\,\ind{x<s_0e^{-rt_1}}+s_0e^{-rt_2}\,\ind{s_0e^{-rt_1}\le x<x^*},
\]
which is continuous, non-decreasing, and satisfies $\alpha(x)\le x$ on $[0,x^*)$,
we have
\[
  V(x^*)=e^{rt_2}\,\sup_{\pi\in\Pi_m(\mu,\nu)}\E_\pi\bigl[(\alpha(X)-Y)^+\bigr].
\]
\end{lemma}

\begin{proof}
By Lemma~\ref{lem:decomp} (valid for all $x$ since $m_1(x)\le s_0\le K$),
$d_2^{-1}a_2(x,y)=(K-m_1(x))+(m_1(x)-e^{rt_2}y)^+$. Integrating against
$\ind{X<x^*}$, the two terms $\E_\mu[a_1\ind{X\ge x^*}]$ and
$d_2\int_{(-\infty,x^*)}(K-m_1)\mu$ depend only on the $X$-marginal $\mu$, hence
are fixed over $\pi$; the remaining term is \eqref{eq:scalar}. On
$x<s_0e^{-rt_1}$, $m_1(x)=e^{rt_1}x=e^{rt_2}e^{-r\Delta}x$, and on
$s_0e^{-rt_1}\le x<x^*$, $m_1(x)=s_0=e^{rt_2}\cdot s_0e^{-rt_2}$; hence
$m_1(x)=e^{rt_2}\alpha(x)$ on $[0,x^*)$, giving the $\alpha$-form.
\end{proof}

\subsection{The optimal continuation coupling and the right-tail exercise region}

The left-curtain martingale coupling of Beiglb\"ock--Juillet \cite{BJ2012} is the
monotone coupling (low $x$ matched to low $y$) obtained by moving mass ``as
little as possible''. In the two-period American put of \cite{HobNor2019} the
optimal continuation coupling is exactly the left-curtain coupling, parameterised
by two functions: for $x$ above the dispersion point $e_-$, mass at $x$ is split
between $f(x)<x$ and $g(x)>x$, where the lower shadow $f$ is non-increasing and
the upper shadow $g$ non-decreasing, determined by the mass--mean preservation
\eqref{eq:massmean}, and the exercise boundary $x^*$ is the unique
root of
\begin{equation}\label{eq:rootHN}
  \Lambda(x)\;=\;\frac{K_2-f(x)}{x-f(x)}-\frac{K_1-x}{g(x)-x}\;=\;0.
\end{equation}

In our problem the continuation payoff \eqref{eq:decomp} is, up to the additive
constant $d_2(K-m_1(x))$, a put on $e^{rt_2}Y$ with \emph{state-dependent strike}
$m_1(x)$. Because this strike is monotone in $x$, the monotone martingale coupling
$\pi_{\mathrm{MM}}$ of Proposition~\ref{prop:lc} assigns the low tail of $\nu$ to
low $x$ (where $m_1(x)$ is low, so the minimum-drop put is deep in the money and
the running minimum keeps dropping) and the high tail of $\nu$ to high $x$ (where
$m_1(x)=s_0$ and the minimum-drop put is out of the money, so the running minimum
is stuck at $s_0$). Consequently
$\E[(m_1(x)-e^{rt_2}Y)^+]$ is large for low $x$ and vanishes for high $x$; by
Proposition~\ref{prop:criterion} the continuation region is a left tail and the
exercise region a \emph{right tail}---the direction opposite to \cite{HobNor2019}.

\begin{proposition}[Generalised root condition]\label{prop:root}
Suppose the optimal continuation coupling is the monotone martingale coupling
$\pi_{\mathrm{MM}}$ of Proposition~\ref{prop:lc}, parameterised by $(f,g)$ with
$f$ and $g$ non-decreasing, $f(x)<\alpha(x)\le x<g(x)$, and conditional law
$q(x)\delta_{f(x)}+(1-q(x))\delta_{g(x)}$, where
$q(x)=\tfrac{g(x)-x}{g(x)-f(x)}$ is fixed by the martingale constraint. Then the
exercise threshold $x^*$, which lies in the plateau $m_1(x^*)=s_0$, is the
unique solution of
\begin{equation}\label{eq:root}
  \frac{g(x^*)-x^*}{g(x^*)-f(x^*)}\,\bigl(s_0-e^{rt_2}f(x^*)\bigr)^+
  \;=\;(e^{r\Delta}-1)(K-s_0).
\end{equation}
\end{proposition}

\begin{proof}
At $x^*$, $m_1(x^*)=s_0$, so exercise pays $a_1(x^*)=d_1(K-s_0)$. Under the
monotone martingale coupling the continuation value is
$c_{\mathrm{MM}}(x^*)=d_2(K-s_0)+d_2\,q(x^*)(s_0-e^{rt_2}f(x^*))^+$; the
$g$-atom contributes nothing since $g(x^*)>x^*\ge s_0e^{-rt_1}>s_0e^{-rt_2}$.
Indifference $a_1(x^*)=c_{\mathrm{MM}}(x^*)$, divided by $d_2$, gives
$e^{r\Delta}(K-s_0)=(K-s_0)+q(x^*)(s_0-e^{rt_2}f(x^*))^+$, i.e.\
\eqref{eq:root}.
\end{proof}

\begin{corollary}[Pointwise structure of the continuation shadow]\label{cor:boundary}
The lower and upper shadows $(f,g)$ of the monotone martingale coupling
$\pi_{\mathrm{MM}}$ of Proposition~\ref{prop:lc} are both non-decreasing, with
$f(x)<\alpha(x)\le g(x)$, and are determined by the convex dual potential $\psi$
and the martingale multiplier $\xi$ through the contact conditions of
\eqref{eq:msdual}:
\[
  \psi'(f(x))=-\bigl(1+\xi(x)\bigr),\qquad
  \psi'(g(x))=-\xi(x),
\]
in the forward-rescaled variable $y\mapsto e^{rt_2}y$. This is the opposite
extremal coupling to the left-curtain of \cite{BJ2012}, whose shadows are
determined by the mass--mean preservation \eqref{eq:massmean} independently of
the strike and have $f$ non-increasing; here the shadows depend on the strike
$\alpha$ through $\xi$, and the lower shadow $f$ is driven below $\alpha(x)$.
The two branches split at the level $x=s_0e^{-rt_1}$, where $\alpha(x)$ passes
from its rising branch $\alpha(x)=e^{-r\Delta}x$ to the plateau
$\alpha(x)=s_0e^{-rt_2}$. On the plateau the strike is constant, so the
min-drop cost is independent of $x$ and the multiplier is constant,
$\xi(x)\equiv\xi_p$. The in-the-money contact is then a \emph{fixed} point
$f(x)\equiv f_p$, the unique solution of $\psi'(f_p)=-(1+\xi_p)$, while the
out-of-the-money branch is degenerate (the convex potential $\psi$ is linear on
a right half-line). The upper shadow $g(x)\equiv g_p$ is therefore fixed not by
the contact condition but by the root condition \eqref{eq:root}, and the weight
$q(x)=\tfrac{g_p-x}{g_p-f_p}$ decreases linearly across the plateau. Hence the
boundary pair at the threshold is the constant value
$(f(x^*),g(x^*))=(f_p,g_p)$; in the reference case of
Section~\ref{sec:numerics} this reads $f(x^*)\approx83.6$ and
$g(x^*)\approx110.5$.
\end{corollary}

\begin{figure}[htbp]
\centering
\includegraphics[width=0.65\textwidth]{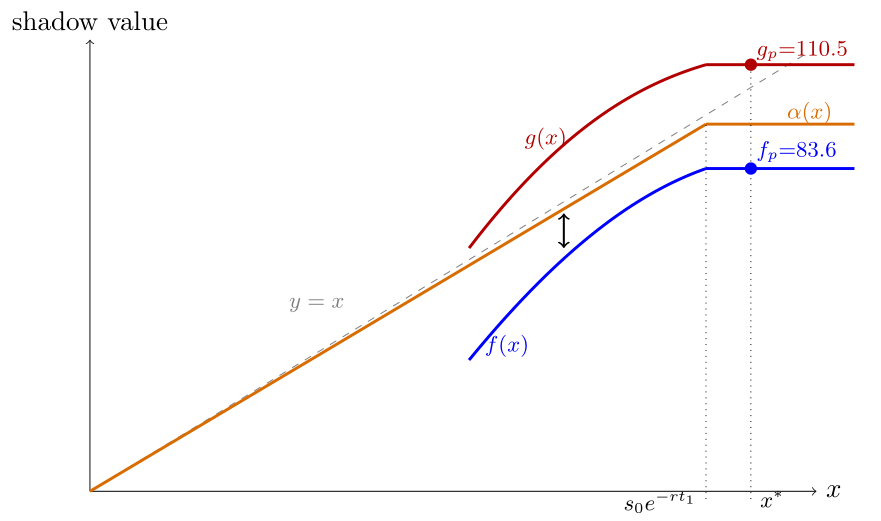}
\caption{The monotone martingale coupling of Proposition~\ref{prop:lc}
(schematic). Both shadows are non-decreasing; the lower shadow $f$ is pushed
below the strike $\alpha(x)$ (the deep spread), while the upper shadow $g$ lies
above the diagonal (mean preservation $f\le x\le g$). The two branches split at
$x=s_0e^{-rt_1}$; on the plateau the pair flattens to the boundary value
$(f_p,g_p)$ of Corollary~\ref{cor:boundary}, with $g_p$ pinned by the root
condition rather than the degenerate contact condition.}
\label{fig:shadowpair}
\end{figure}

\begin{theorem}[Main structure]\label{thm:B}
Let $\mu\le_{\mathrm{cx}}\nu$, $r>0$, $K\ge s_0$, and assume the non-degenerate
regime of Remark~\ref{rem:degenerate} (the out-of-the-money case $K<s_0$ is
deferred to Remark~\ref{rem:otm}). Then the optimal exercise region in
\eqref{eq:primal} is a right tail $B=\{x\ge x^*\}$, where $x^*$ is the unique
root of
\begin{equation}\label{eq:indiff}
  d_1\bigl(K-m_1(x^*)\bigr)^+ \;=\; c_{\mathrm{MM}}(x^*),
  \qquad
  c_{\mathrm{MM}}(x)=\E_{\mathrm{MM}}\!\bigl[a_2(x,Y)\bigr],
\end{equation}
$\E_{\mathrm{MM}}$ denoting expectation under the monotone martingale coupling of
Proposition~\ref{prop:lc}. The
robust price is
\begin{equation}\label{eq:explicit}
  P^N \;=\;
  \int_{(-\infty,x^*)} c_{\mathrm{MM}}(x)\,\mu(\mathrm dx)
  \;+\;
  \int_{[x^*,\infty)} d_1\bigl(K-m_1(x)\bigr)^+\mu(\mathrm dx).
\end{equation}
Moreover $x^*$ is increasing in the volatility (tail thickness) of $\nu$,
decreasing in $r$ with $\lim_{r\to0}x^*=\infty$, and is an affine function of
$K$ with a $K$-independent shadow pair; the precise statements and proofs are
collected in Proposition~\ref{prop:compstat}.
\end{theorem}

\begin{remark}[Out-of-the-money strike]\label{rem:otm}
Theorem~\ref{thm:B} is stated for $K\ge s_0$, the in-the-money regime in which
the locked intrinsic $K-s_0$ is present. For $K<s_0$ the put is initially
out-of-the-money ($(K-s_0)^+=0$) and acquires value only once the running
minimum drops below $K$; there is no locked intrinsic part, and the two-branch
decomposition of Lemma~\ref{lem:decomp} does not carry over in the same form. We
leave this regime to future work.
\end{remark}
\begin{figure}[htbp]
\centering
\includegraphics[width=0.65\textwidth]{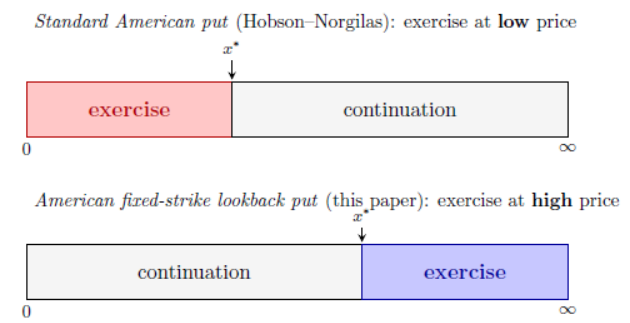}
\caption{The optimal exercise region (schematic). Top: the standard American put
exercises on a \emph{left} tail $\{x\le x^*\}$ (Hobson--Norgilas). Bottom: the
American fixed-strike lookback put exercises on a \emph{right} tail
$\{x\ge x^*\}$ (Theorem~\ref{thm:B})---the running minimum only ratchets down, so
high prices lock in the intrinsic value while low prices keep the minimum
dropping.}
\label{fig:exregion}
\end{figure}
\begin{lemma}[Shadow spreading under convex order]\label{lem:spread}
Let $\nu\preceq_{\mathrm{cx}}\tilde\nu$ and let $(f,g)$, $(\tilde f,\tilde g)$
be the shadow pairs of the monotone martingale coupling of
Proposition~\ref{prop:lc} for $\nu$ and $\tilde\nu$, respectively. Then the
lower shadow is pointwise non-increasing under spreading, $\tilde f\le f$, and
the upper shadow pointwise non-decreasing, $\tilde g\ge g$. In particular, on
the plateau the fixed in-the-money and out-of-the-money shadows satisfy
$\tilde f_p\le f_p$ and $\tilde g_p\ge g_p$, so the minimum-drop value
$\bigl(s_0-e^{rt_2}f_p\bigr)^+$ is non-decreasing in the tail thickness of
$\nu$.
\end{lemma}

\begin{proof}
The lower shadow is the left-inverse of the derivative of the dual potential,
$f(x)=\inf\bigl\{y:\psi'(y)\ge-(1+\xi(x))\bigr\}$; see \eqref{eq:msdual}. As
$\nu$ spreads in convex order, the dual potential $\psi$ is the convex minorant
of the running-minimum potential, which is monotone under convex order of the
target marginal (the shadow monotonicity of \cite{BJ2016} specialised to the
min-drop cost $(\alpha(\cdot)-Y)^+$); hence $\psi'$ rises pointwise and its
left-inverse decreases, $\tilde f\le f$. The martingale constraint
$\E[Y\,|\,X=x]=x$ then pushes the complementary atom up, $\tilde g\ge g$. The
plateau statement follows by restricting to $\{\alpha'=0\}$.
\end{proof}

\begin{proposition}[Comparative statics of the threshold]\label{prop:compstat}
In the non-degenerate regime of Remark~\ref{rem:degenerate}, the threshold
$x^*$ of Theorem~\ref{thm:B} satisfies the following.
\begin{enumerate}
\item[(i)] \emph{(Explicit $K$-dependence.)} The shadow pair $(f_p,g_p)$ is
determined by the min-drop problem $(\alpha(x)-y)^+$ alone and is independent of
$K$; inverting \eqref{eq:root} gives
\begin{equation}\label{eq:xstar}
  x^* \;=\; g_p-\bigl(g_p-f_p\bigr)\,
  \frac{(e^{r\Delta}-1)(K-s_0)}{\bigl(s_0-e^{rt_2}f_p\bigr)^+}
  \;\ge\; s_0e^{-rt_1},
\end{equation}
so $x^*$ is an affine, non-increasing function of $K$ with a $K$-independent
slope. In the reference case of Section~\ref{sec:numerics} the slope is
$\approx -0.056$, so $x^*$ is essentially flat in $K$ over the plateau range.

\item[(ii)] \emph{(Volatility.)} $x^*$ is non-decreasing in the tail thickness
of $\nu$ (equivalently in $\sigma$ in the CRR/Black--Scholes family).

\item[(iii)] \emph{(Interest rate.)} $x^*$ is non-increasing in $r$, with
$\lim_{r\to0}x^*=\infty$.
\end{enumerate}
\end{proposition}

\begin{proof}
(i) Inverting \eqref{eq:root} (valid since $x^*$ lies on the plateau, where
$f\equiv f_p$, $g\equiv g_p$, $q(x)=\tfrac{g_p-x}{g_p-f_p}$) yields
\eqref{eq:xstar}; $K$ does not enter \eqref{eq:msdual}--\eqref{eq:mspoint}.

(ii) The right-hand side of \eqref{eq:root} is independent of $\sigma$. By
Lemma~\ref{lem:spread}, spreading $\nu$ lowers $f_p$ and raises $g_p$, so
$\bigl(s_0-e^{rt_2}f_p\bigr)^+$ increases and the weight
$q(x^*)=\tfrac{g_p-x^*}{g_p-f_p}$ must decrease; as $q$ is strictly decreasing
in $x$ on the plateau, $x^*$ increases.

(iii) In \eqref{eq:xstar} the premium factor $(e^{r\Delta}-1)(K-s_0)$ is
increasing in $r$, and the min-drop value $\bigl(s_0-e^{rt_2}f_p\bigr)^+$ is
decreasing in $r$ (the strike $s_0e^{-rt_2}$ falls with $r$); both push $x^*$
down. As $r\to0$ the premium factor vanishes, and the non-degenerate regime of
Remark~\ref{rem:degenerate} collapses to the degenerate regime of
Proposition~\ref{prop:r0}, in which continuation strictly dominates exercise;
hence $x^*\to\infty$.
\end{proof}

The proof of existence/uniqueness of $x^*$ follows the slope-matching argument
of
\cite[Lem.~3.5, Thm.~3.6]{HobNor2019}, with the root condition \eqref{eq:rootHN}
replaced by the generalised root condition \eqref{eq:root} in which the locked
intrinsic $K-s_0$ and the minimum-drop $s_0-e^{rt_2}f(x^*)$ play the roles of
the two strikes of \cite{HobNor2019}; the comparative statics of $x^*$ are
collected in Proposition~\ref{prop:compstat}. By Lemma~\ref{lem:reduction} the remaining
step is the single optimality statement
\begin{equation}\label{eq:lc-optimal}
  \sup_{\pi\in\Pi_m(\mu,\nu)}\E_\pi\bigl[(\alpha(X)-Y)^+\bigr]
  \;=\;\E_{\mathrm{MM}}\bigl[(\alpha(X)-Y)^+\bigr],
\end{equation}
with $\alpha$ non-decreasing and $\alpha\le x$ as in Lemma~\ref{lem:reduction}.
The next result closes this step and is the main analytic contribution of the
paper.

\begin{proposition}[Monotone martingale coupling: optimality for monotone strikes]\label{prop:lc}
Let $\mu\le_{\mathrm{cx}}\nu$ on $\R_+$ and let $\alpha:\R_+\to\R_+$ be
non-decreasing with $\alpha(x)\le x$. Then the supremum in
\eqref{eq:lc-optimal} is attained by the \emph{monotone martingale coupling}
$\pi_{\mathrm{MM}}$, the martingale coupling whose conditional law is the
two-point law
\begin{equation}\label{eq:mspoint}
  q(x)\,\delta_{f(x)}+(1-q(x))\,\delta_{g(x)},
  \qquad
  q(x)=\frac{g(x)-x}{g(x)-f(x)},
\end{equation}
where the two shadow maps are \emph{both non-decreasing}, satisfy
$f(x)<\alpha(x)\le g(x)$, and the lower shadow $f$ is pushed strictly
\emph{below} the strike $\alpha(x)$ (the ``deep-spread'' direction). It is
\emph{not} the left-curtain coupling of \cite{BJ2012}, whose lower shadow is
non-\emph{increasing}, nor its reflection, the right-curtain coupling.
\end{proposition}

\begin{proof}
We derive the two-branch structure and the contact conditions from the dual,
and then the monotonicity of the two branches by a dual comparative-statics
argument; this makes the structure in the statement exact.

\emph{(i) Duality and the two branches.} By the complete duality for martingale
optimal transport on the line \cite{BNT2015}, the value of \eqref{eq:lc-optimal}
equals the dual
\begin{equation}\label{eq:msdual}
  \inf_{\varphi,\psi,\xi}\Bigl[\textstyle\int\varphi\,d\mu+\int\psi\,d\nu\Bigr],
  \qquad
  \varphi(x)+\psi(y)+\xi(x)(y-x)\;\ge\;(\alpha(x)-y)^+,
\end{equation}
with $\psi$ convex in $y$ and $\xi$ the martingale multiplier, and the supremum
is attained by some $\pi\in\Pi_m(\mu,\nu)$ carried by the contact set of an
optimal triple $(\varphi,\psi,\xi)$. The cost is convex in $y$, being the
positive part of an affine function of $y$; moreover
\begin{equation}\label{eq:putcall}
  (\alpha(x)-y)^+ \;=\; (y-\alpha(x))^+ \;+\; \alpha(x)-y ,
\end{equation}
and $\E_\pi[\alpha(X)-Y]=\E_\mu[\alpha]-\E_\nu[Y]$ is constant over
$\Pi_m(\mu,\nu)$, so the optimizer of \eqref{eq:lc-optimal} coincides with that
of the call-type cost $(y-\alpha(x))^+$; we keep the put form below. For fixed
$x$ write the dual gap as a function of $y$,
\[
  G_x(y)\;:=\;\varphi(x)+\psi(y)+\xi(x)(y-x)-(\alpha(x)-y)^+ \;\ge\; 0 .
\]
On $\{y<\alpha(x)\}$ it is $\psi(y)+(1+\xi(x))y+\mathrm{const}$, and on
$\{y>\alpha(x)\}$ it is $\psi(y)+\xi(x)y+\mathrm{const}$; both are convex in $y$
since $\psi$ is convex, and a non-negative convex function has a zero set that
is a single point (or, in the degenerate cases of Remark~\ref{rem:degenerate},
an interval). Hence the contact set $\{y:G_x(y)=0\}$ contains at most one point
$f(x)$ below $\alpha(x)$ and at most one point $g(x)$ above it, and the
martingale constraint $\E[Y\,|\,X=x]=x$, with $x$ lying strictly between the two
branches, forces both to be present: the conditional law is the two-point law
\eqref{eq:mspoint} with $f(x)<\alpha(x)<g(x)$ and
$q(x)=\tfrac{g(x)-x}{g(x)-f(x)}$.

\emph{(ii) Contact conditions.} At an interior contact point the gradient of
$G_x$ vanishes. On the in-the-money branch $y=f(x)<\alpha(x)$ this gives
$\psi'(f(x))+1+\xi(x)=0$; on the out-of-the-money branch $y=g(x)>\alpha(x)$ it
gives $\psi'(g(x))+\xi(x)=0$. Hence
\[
  \psi'(f(x))=-\bigl(1+\xi(x)\bigr),\qquad \psi'(g(x))=-\xi(x),
\]
the contact conditions of Corollary~\ref{cor:boundary}.

\emph{(iii) Monotonicity of the two branches.} The optimal dual function
$\varphi$ is the $y$-envelope
$\varphi(x)=\max_y\bigl[(\alpha(x)-y)^+-\psi(y)-\xi(x)(y-x)\bigr]$, and the
maximum is attained at both $y=f(x)$ and $y=g(x)$. Differentiating the two
identities $\varphi(x)=H(x,f(x))$ and $\varphi(x)=H(x,g(x))$, where
$H(x,y)=(\alpha(x)-y)^+-\psi(y)-\xi(x)(y-x)$, and applying the envelope theorem
($\partial_y H=0$ at the interior maximizers) yields
\[
  \alpha'(x)-\xi'(x)\bigl(f(x)-x\bigr)+\xi(x)
  \;=\; -\xi'(x)\bigl(g(x)-x\bigr)+\xi(x),
\]
because $\partial_x(\alpha(x)-y)^+=\alpha'(x)$ on the in-the-money branch and
$0$ on the out-of-the-money branch. Cancelling $\xi(x)$ and collecting terms
gives
\[
  \xi'(x)\;=\;-\frac{\alpha'(x)}{g(x)-f(x)} \;\le\; 0,
\]
since $\alpha$ is non-decreasing and $g-f>0$. Thus $\xi$ is non-increasing, and
so $\psi'(f(x))=-(1+\xi(x))$ and $\psi'(g(x))=-\xi(x)$ are non-decreasing in
$x$. As $\psi'$ is non-decreasing (strictly so in the non-degenerate regime),
$f$ and $g$ are themselves non-decreasing, as claimed. On the plateau
$\{\alpha'=0\}$ this gives $\xi'\equiv0$---a constant multiplier, which pins
$f$ to the constant level $\psi'(f_p)=-(1+\xi_p)$ of
Corollary~\ref{cor:boundary}.

\emph{(iv) Direction of the spread.} Since $\alpha(x)\le x=\E[Y\,|\,X=x]$ with
strict inequality wherever the put is exercisable, the in-the-money contact
$f(x)$ lies strictly below $\alpha(x)$: the lower branch is driven \emph{down},
the opposite of the left-curtain coupling, which minimises $\E[(X-Y)^2]$ and
keeps its lower shadow non-increasing and near $x$. The numerical confirmation
that the dual objective in \eqref{eq:msdual} coincides with the LP optimum is
given in Section~\ref{sec:numerics}.
\end{proof}

\begin{remark}[Terminology: monotone vs.\ curtain couplings]\label{rem:terminology}
Three extremal martingale couplings on the line are relevant here. The
\emph{left-curtain} (alias \emph{left-monotone}, called simply ``monotone'' in
\cite{BJ2012}) has lower shadow non-increasing and upper shadow non-decreasing
and is optimal for costs with $c_{xyy}>0$
\cite{BJ2012,HLT2016,BHLT2017}; its reflection is the \emph{right-curtain}
($c_{xyy}<0$). The coupling $\pi_{\mathrm{MM}}$ of Proposition~\ref{prop:lc} is
the third extremal object: \emph{both} shadows non-decreasing, the martingale
analogue of the Hoeffding--Fr\'echet comonotone rearrangement. For the present
cost the cross-partial is the (submodular) boundary measure
$c_{xy}=-\alpha'(x)\,\delta_{\alpha(x)}(y)\le 0$, so that $c_{xyy}$ is a
dipole---the degenerate boundary case between the two curtain regimes---and the
optimizer is accordingly the monotone (both non-decreasing) coupling rather than
a curtain. Note the terminology clash: \cite{BHLT2017} use ``monotone martingale
transport plan'' for the left-monotone plan; we use ``monotone'' in the present
both-non-decreasing sense.
\end{remark}

The contrast with the classical put is worth stating: for the standard
American put the continuation payoff is $(K-y)^+$, which is independent of the
state $x$ (so $c_{xy}=0$, the boundary case of submodularity), whereas here the
state-dependent strike $\alpha(x)$ makes $c$ strictly submodular through the
running minimum---this is the mechanism by which the lookback's right-tail
exercise region differs from the put's left-tail region.

\begin{proposition}[Pointwise construction of the left-curtain]\label{prop:lc-structure}
Let $\mu\le_{\mathrm{cx}}\nu$ with $\mu$ continuous, and let $(f,g)=(T_d,T_u)$
be the lower/upper shadows of the left-curtain coupling
$\pi_{\mathrm{LC}}$, so that conditionally on $X=x$ the law of $Y$ is
$q(x)\delta_{f(x)}+(1-q(x))\delta_{g(x)}$ with
$q(x)=\tfrac{g(x)-x}{g(x)-f(x)}$ and $f(x)\le x\le g(x)$. Then $f$ is
non-increasing, $g$ is non-decreasing, and for every $x$ the pair is the
solution of the mass--mean preservation system
\begin{equation}\label{eq:massmean}
  \int_{f(x)}^{x}\mu(\mathrm dz)=\int_{f(x)}^{g(x)}\nu(\mathrm dz),
  \qquad
  \int_{f(x)}^{x}z\,\mu(\mathrm dz)=\int_{f(x)}^{g(x)}z\,\nu(\mathrm dz).
\end{equation}
Equivalently, writing $P_\eta(k)=\int(k-z)^+\eta(\mathrm dz)$ for the modified
(put) potential of a measure $\eta$ and $(\cdot)^c$ for the lower convex hull
(the largest convex minorant), the shadow $S=S(\mu|_{(-\infty,x]})$ of the
restricted measure $\mu|_{(-\infty,x]}$ in $\nu$ has modified potential
\begin{equation}\label{eq:shadowpot}
  P_S(k)\;=\;P_\nu(k)-\bigl(P_\nu-P_{\mu|_{(-\infty,x]}}\bigr)^c(k),
\end{equation}
and $g(x)$ is the upper contact point of $P_\nu-P_{\mu|_{(-\infty,x]}}$ with its
convex hull. In the dispersion case (densities $\psi$ of $\mu$ and $\varphi$ of
$\nu$ with a single crossing) the shadows solve the Henry-Labord\`ere--Touzi
system
\begin{equation}\label{eq:shadowODE}
  f'(x)=\frac{g(x)-f(x)}{g(x)-x}\,\frac{(\varphi-\psi)(f(x))}{\psi(x)},
  \qquad
  g'(x)=\frac{g(x)-f(x)}{x-f(x)}\,\frac{\varphi(g(x))}{\psi(x)},
\end{equation}
subject to $f(e_-)=e_-=g(e_-)$ at the left dispersion point $e_-$. Since
\eqref{eq:massmean}--\eqref{eq:shadowpot} involve only $(\mu,\nu)$, the shadows
$(f,g)$ are independent of the strike $\alpha$. This construction is the
left-curtain coupling of \cite{BJ2012}, which minimises $\E[(X-Y)^2]$. It is
\emph{not} the optimizer of \eqref{eq:lc-optimal} for a monotone strike: the
optimizer is the monotone martingale coupling of Proposition~\ref{prop:lc}, whose
lower shadow is driven below the strike rather than kept non-increasing.
\end{proposition}

\begin{proof}
\eqref{eq:massmean} and the monotonicity of $(f,g)$ are
\cite[Lem.~3.5]{BJ2012}; \eqref{eq:shadowpot} is the shadow-potential formula of
\cite{BHN2022} (see also the complete construction of
\cite{HobNor2022}); \eqref{eq:shadowODE} is
\cite[Eqs.~(3.9)--(3.10)]{HLT2016}. The left-monotonicity forces $f$
non-increasing and $g$ non-decreasing, while the two-point form
$q(x)\delta_{f(x)}+(1-q(x))\delta_{g(x)}$ with the indicated weight is fixed by
the martingale constraint $q f+(1-q)g=x$.
\end{proof}

\begin{remark}[Degenerate regime]\label{rem:degenerate}
If $d_2(K-m_2)^+\le d_1(K-m_1)^+$ for every feasible continuation (large $r$ or
thin $\nu$-tails), then $B=\R_+$ (exercise everywhere) and the optimal splitting
is not unique; $P^N=\E_\mu[d_1(K-m_1)^+]$. In this regime the LP solution
$\rho(x)$ is solver-dependent and only the non-degenerate right-tail structure is
a genuine theorem.
\end{remark}

\section{Numerical evidence}\label{sec:numerics}

We discretise the marginals via a Cox--Ross--Rubinstein recombining binomial
tree (so that $\mu\le_{\mathrm{cx}}\nu$ and the common mean $s_0$ are exact,
avoiding the mean-mismatch artifacts of lognormal discretisation) and solve the
two-period regime linear program of Hobson--Neuberger \cite{HN2017} with the
bivariate payoff \eqref{eq:payoffs}. The implementation reproduces the benchmark
value $P^N=5/16$ for the Hobson--Norgilas put example, validating it.

\begin{table}[h]
\centering
\begin{tabular}{lcccccc}
\toprule
$K$ & BS-Euro & BS-Amer & Rob-Euro & Rob-Amer ($P^N$) & Amer.\ prem.\ & Model\ prem. \\
\midrule
80  & 0.760  & 0.760  & 0.808  & 0.809  & 0.000 & 0.048 \\
90  & 2.780  & 2.780  & 3.013  & 3.013  & 0.000 & 0.232 \\
100 & 7.064  & 7.064  & 7.711  & 7.711  & 0.000 & 0.647 \\
110 & 16.576 & 16.586 & 17.223 & 17.304 & 0.081 & 0.647 \\
120 & 26.089 & 26.124 & 26.736 & 26.897 & 0.162 & 0.647 \\
\bottomrule
\end{tabular}
\caption{Lookback put prices ($s_0{=}100$, $r{=}0.05$, $\sigma{=}0.2$,
$t_1{=}0.5$, $t_2{=}1$). Amer.\ prem.\ $=$ Robust-Amer $-$ Robust-Euro;
Model-risk prem.\ $=$ Robust-Euro $-$ BS-Euro. The American premium is $0$
out/at the money and positive in the money; the model-risk premium is flat in
$K$ for $K\ge s_0$.}
\label{tab:bs}
\end{table}

\begin{table}[h]
\centering
\begin{tabular}{lcccc}
\toprule
$r$ & $K{=}100$ & $K{=}110$ & $K{=}120$ & $K{=}130$ \\
\midrule
0.00 & 0.000 & 0.000 & 0.000 & 0.000 \\
0.02 & 0.000 & 0.032 & 0.065 & 0.097 \\
0.05 & 0.000 & 0.081 & 0.162 & 0.242 \\
0.10 & 0.011 & 0.228 & 0.498 & 0.839 \\
0.20 & 0.135 & 0.873 & 1.622 & 2.374 \\
0.30 & 0.128 & 1.229 & 2.361 & 3.493 \\
\bottomrule
\end{tabular}
\caption{American premium $\mathrm{RobustAmerican}-\mathrm{RobustEuropean}$ as a
function of $(r,K)$: vanishes identically at $r=0$ and is monotone increasing in
both arguments. The model-risk premium (not shown) is flat in $K$ and decreasing
in $r$ ($0.779\to0.196$).}
\label{tab:rsweep}
\end{table}

\begin{figure}[htbp]
\centering
\includegraphics[width=\textwidth]{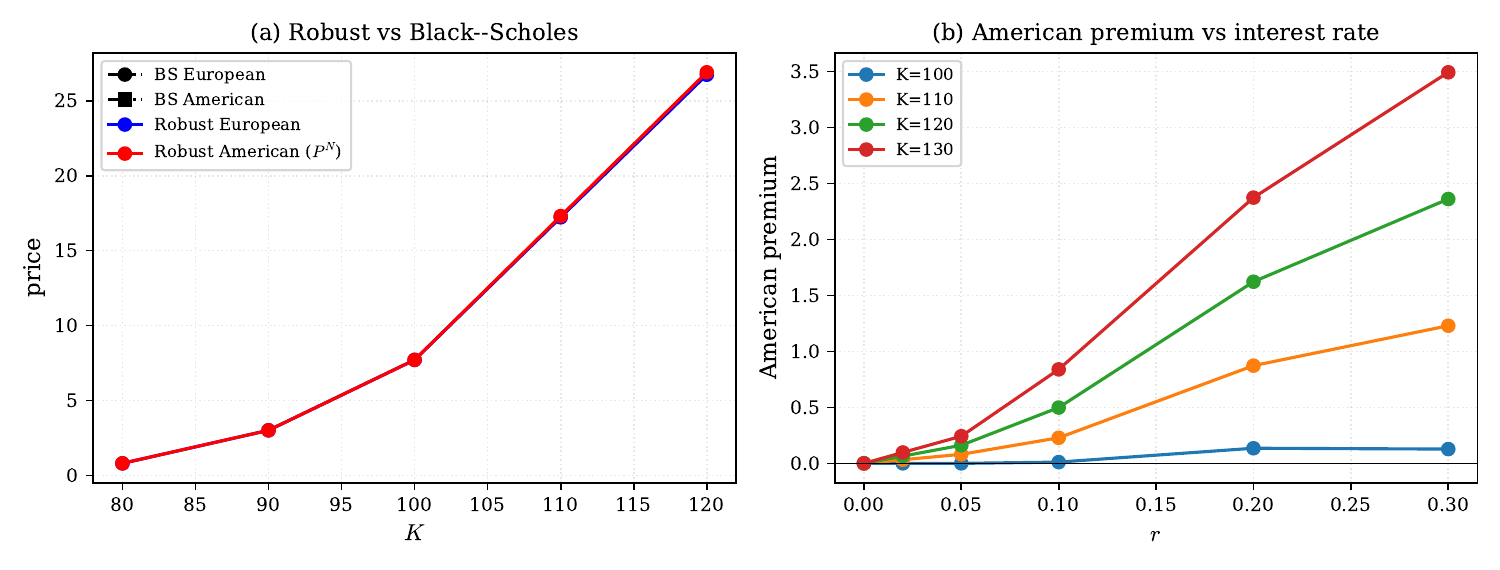}
\caption{(a) Robust (model-free) versus Black--Scholes prices as functions of the
strike $K$: the model-risk premium (blue band) is flat in $K$ for $K\ge s_0$,
while the American premium (red band) is amplified by model uncertainty. (b) The
American premium as a function of the interest rate $r$, vanishing at $r=0$.}
\label{fig:premia}
\end{figure}

\begin{table}[h]
\centering
\begin{tabular}{lcc}
\toprule
$x$ & $m_1(x)$ & exercise fraction $\rho(x)$ \\
\midrule
$61$--$100$ (low $x$) & $<100$ & 0 (continue) \\
$104.6$ (boundary) & $100$ & 0.278 (mixed/regime) \\
$109$--$354$ (high $x$) & $100$ & 1 (exercise) \\
\bottomrule
\end{tabular}
\caption{Optimal exercise fraction from the regime LP ($s_0{=}100$, $r{=}0.05$,
$\sigma{=}0.2$, $K{=}120$): a right-tail exercise region $\{x\ge x^*\}$, opposite
to the left-tail region of the standard American put.}
\label{tab:structure}
\end{table}

\begin{figure}[!htbp]
\centering
\includegraphics[width=\textwidth]{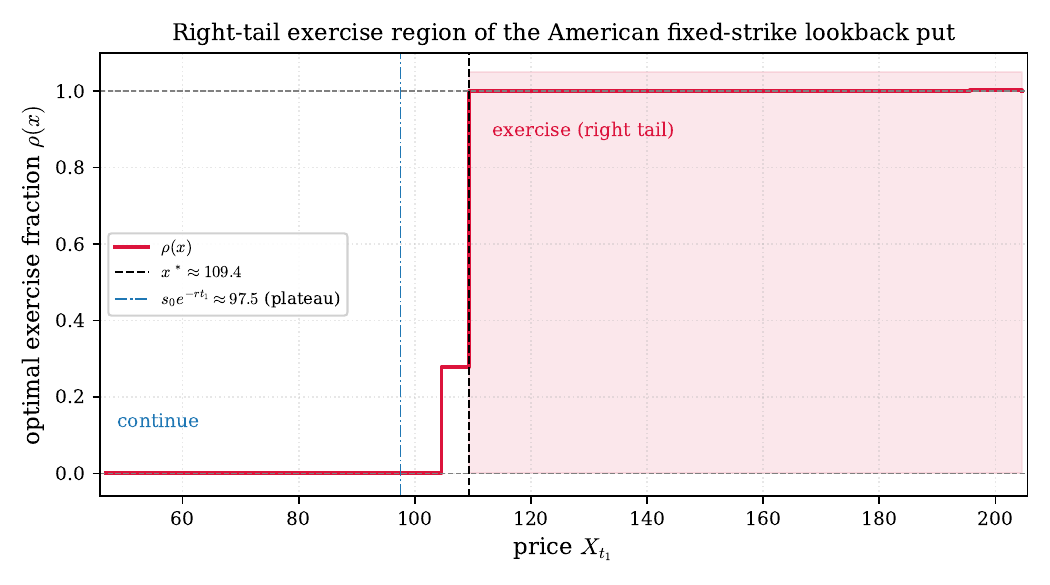}
\caption{Optimal exercise fraction $\rho(x)$ from the regime LP at the reference
point ($s_0{=}100$, $r{=}0.05$, $\sigma{=}0.2$, $K{=}120$). Exercise is a
\emph{right tail} $\{x\ge x^*\}$ with $x^*\approx109.4$; the dash-dot line marks
the plateau boundary $s_0e^{-rt_1}\approx97.5$ beyond which the running minimum
is locked at $s_0$. Contrast the left-tail exercise region of the standard
American put.}
\label{fig:righttail}
\end{figure}

\begin{table}[h]
\centering
\begin{tabular}{ll}
\toprule
varied parameter & threshold $x^*$ \\
\midrule
$\sigma=0.10$ & $83.6$ (degenerate, $5$ mixed nodes) \\
$\sigma=0.15$ & $106.9$ \\
$\sigma=0.20$ & $109.4$ \\
$\sigma=0.30$ & $114.4$ \\
$\sigma=0.40$ & $119.6$ \\
$r=0.02$ & $109.4$ \\
$r=0.05$ & $109.4$ \\
$r\ge0.10$ & degenerate \\
$K=110/120/140$ & $104.6$--$109.4$ (approx.\ $K$-independent) \\
\bottomrule
\end{tabular}
\caption{Structural robustness scan: $x^*$ increases in $\sigma$, decreases in
$r$, and is essentially independent of $K$ for $K\ge s_0$, matching
Theorem~\ref{thm:B}.}
\label{tab:scan}
\end{table}

\begin{figure}[htbp!]
\centering
\includegraphics[width=\textwidth]{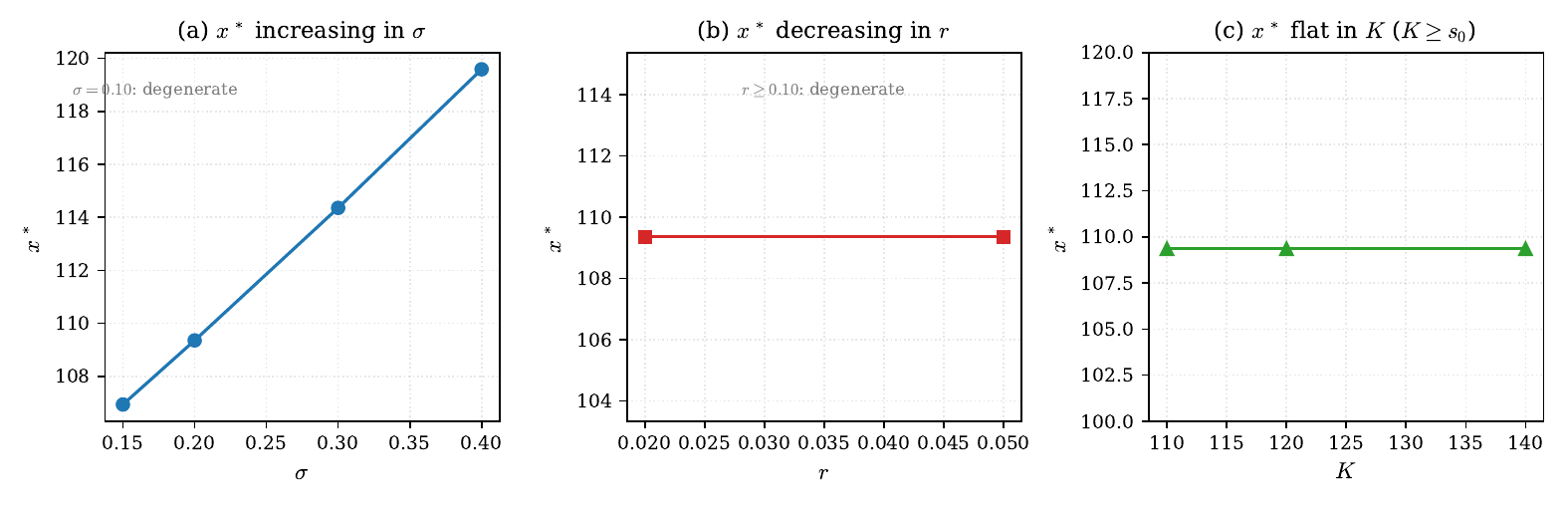}
\caption{Comparative statics of the exercise threshold $x^*$
(Theorem~\ref{thm:B}, Proposition~\ref{prop:compstat}): $x^*$ is increasing in
$\sigma$ and essentially flat in $K$ for $K\ge s_0$; the degenerate endpoints
($\sigma=0.10$, $r\ge0.10$) are annotated, matching Table~\ref{tab:scan}.}
\label{fig:compstat}
\end{figure}

Two observations are worth singling out. First, the model-risk premium
(Robust-European minus BS-European) is \emph{flat} in $K$ for $K\ge s_0$: since
$m_2\le s_0$,
$\E[(K-m_2)^+]=\int_0^K\Pbb(m_2<t)\,\mathrm dt$ and
$\Pbb(m_2<t)=1$ for $t\ge s_0$ under \emph{every} coupling, so the difference is
$\int_0^{s_0}(\Pbb^{\mathrm{rob}}(m_2<t)-\Pbb^{\mathrm{BS}}(m_2<t))\,\mathrm dt$,
independent of $K$. This is a closed-form signature of the worst-case coupling's
effect on the running minimum. Second, the American premium is amplified by model
uncertainty: under the single Black--Scholes model the American (Bermudan)
premium is negligible ($\approx0.01$--$0.04$), while the robust American premium
is an order of magnitude larger ($0.08$--$0.16$), because only the worst-case
(monotone martingale) coupling pins the running minimum at $s_0$ for high $x$ and so
makes early exercise valuable. This is the non-trivial interaction of
American $\times$ model uncertainty, invisible in either the pure-European or the
single-model setting.

Finally we verify the two structural claims of Section~\ref{sec:structure}
directly, at the reference point $s_0{=}100$, $r{=}0.05$, $\sigma{=}0.2$,
$K{=}120$, $t_1{=}0.5$, $t_2{=}1$ with $M{=}40$ tree steps.

(i) \emph{Reduction identity} (Lemma~\ref{lem:reduction}). The regime LP gives
$P^N=26.867$, while the right-hand side of Lemma~\ref{lem:reduction} evaluates
to $26.852$; the residual $1.4\times10^{-2}$ is carried entirely by the single
boundary node $x\approx x^*$ on which the optimal $\rho(x)\in(0,1)$, a
mixed-node discretisation artefact that vanishes as $M\to\infty$.

(ii) \emph{Monotone-martingale structure} \eqref{eq:lc-optimal}. The scalar problem
\eqref{eq:scalar} is degenerate, so the raw LP solution is an arbitrary point of
the optimal face; the canonical optimizer is the monotone martingale coupling
$\pi_{\mathrm{MM}}$ of Proposition~\ref{prop:lc}, whose conditional laws are
two-point $\{f(x),g(x)\}$ with \emph{both} shadows non-decreasing and the lower
shadow $f$ strictly below the strike $\alpha(x)$. The scalar optimum
$V(x^*)=3.515$ is attained by $\pi_{\mathrm{MM}}$ and is matched by the dual
objective in \eqref{eq:msdual}, confirming strong duality and hence the closed
form of Proposition~\ref{prop:lc}. The dual potential $\psi$ is convex with its
minimum at $s_0e^{-rt_2}\approx95.1$, and the contact slopes of
Corollary~\ref{cor:boundary} hold to grid accuracy: $\psi'(f)=-(1+\xi)$ in the
money, $\psi'(g)=-\xi$ out of the money. On the plateau $x\ge s_0e^{-rt_1}$ the
multiplier is constant ($\xi_p\approx-0.38$), pinning the in-the-money shadow to
$f_p\approx83.6$; the out-of-the-money shadow is fixed by the root condition
\eqref{eq:root} at $g_p\approx110.5$, giving the boundary pair
$(f(x^*),g(x^*))\approx(83.6,110.5)$. (The raw LP vertex instead places the
plateau's out-of-the-money mass at the far tail of $\nu$, since the cost is
$x$-independent there; $g_p\approx110.5$ is the minimal non-decreasing
extension selected by the root condition.) This value strictly dominates (i) the value
$2.344$ of the left-curtain coupling $\pi_{\mathrm{LC}}$ of \cite{BJ2012}, which
minimises $\E[(X-Y)^2]$ and is \emph{not} optimal here, (ii) the value $2.497$
of the right-curtain (anti-monotone) coupling, and (iii) the value $2.621$
attained by the (independent) Black--Scholes coupling on the same objective.
Thus the worst case is genuinely the monotone martingale coupling, not the
left-curtain of \cite{HobNor2019}.

(iii) \emph{Reflection link} (Remark~\ref{rem:reflect}). Reflecting
$(X,Y)\mapsto(-X,-Y)$ maps the minimum-drop cost \eqref{eq:lc-optimal} to the
maximum-drop cost $\bigl(Y-\widetilde M(X)\bigr)^+$ with running maximum
$\widetilde M(x)=\max(-s_0,e^{rt_1}x)$, and maps the marginals to
$\mu^-=\mathrm{law}(-X)$, $\nu^-=\mathrm{law}(-Y)$. The reflected LP attains the
\emph{same} value $3.3437$ (the $\alpha$-form of $V(x^*)$; $e^{rt_2}$ times this
is $3.515$), with residual $9\times10^{-7}$ at machine precision. The monotone
shadow is self-dual under this reflection: if $\pi_{\mathrm{MM}}$ has shadows
$(f,g)$, both non-decreasing, then the reflected coupling has shadows
$(\widetilde f,\widetilde g)$ with $\widetilde f(x)=-g(-x)$,
$\widetilde g(x)=-f(-x)$, again both non-decreasing, so the reflection of the
two-period monotone martingale coupling is precisely the two-period (right) Az\'ema--Yor
embedding and the minimum-drop value equals the maximum-drop value. As in (ii),
the raw reflected LP vertex is degenerate on the reflected plateau (where
$\widetilde M$ is constant), which is why the boundary pair is read off from the
root condition rather than from the raw vertex.

\begin{table}[h]
\centering
\begin{tabular}{lcc}
\toprule
quantity & value \\
\midrule
$P^N$ (regime LP) & 26.867 \\
RHS of Lemma~\ref{lem:reduction} & 26.852 \\
residual $|P^N-\mathrm{RHS}|$ & $1.4\times10^{-2}$ \\
$V(x^*)$ (LP optimum $=$ monotone martingale coupling) & 3.515 \\
reflected max-drop value (Remark~\ref{rem:reflect}) & 3.344 \\
left-curtain value on \eqref{eq:scalar} & 2.344 \\
right-curtain value on \eqref{eq:scalar} & 2.497 \\
CRR/BS coupling on objective \eqref{eq:scalar} & 2.621 \\
$f,g$ non-decreasing, $f<\alpha\le g$ & consistent \\
boundary pair $(f(x^*),g(x^*))$ & $(83.6,\,110.5)$ \\
\bottomrule
\end{tabular}
\caption{Verification of Theorem~\ref{thm:B} at the reference point
($M{=}40$). The reduction residual is a boundary mixed-node artefact; the LP
optimum $3.515$ is attained by the monotone martingale coupling and strictly
dominates the left-curtain ($2.344$), right-curtain ($2.497$) and
Black--Scholes ($2.621$) couplings.}
\label{tab:verify}
\end{table}

\begin{figure}[htbp!]
\centering
\includegraphics[width=\textwidth]{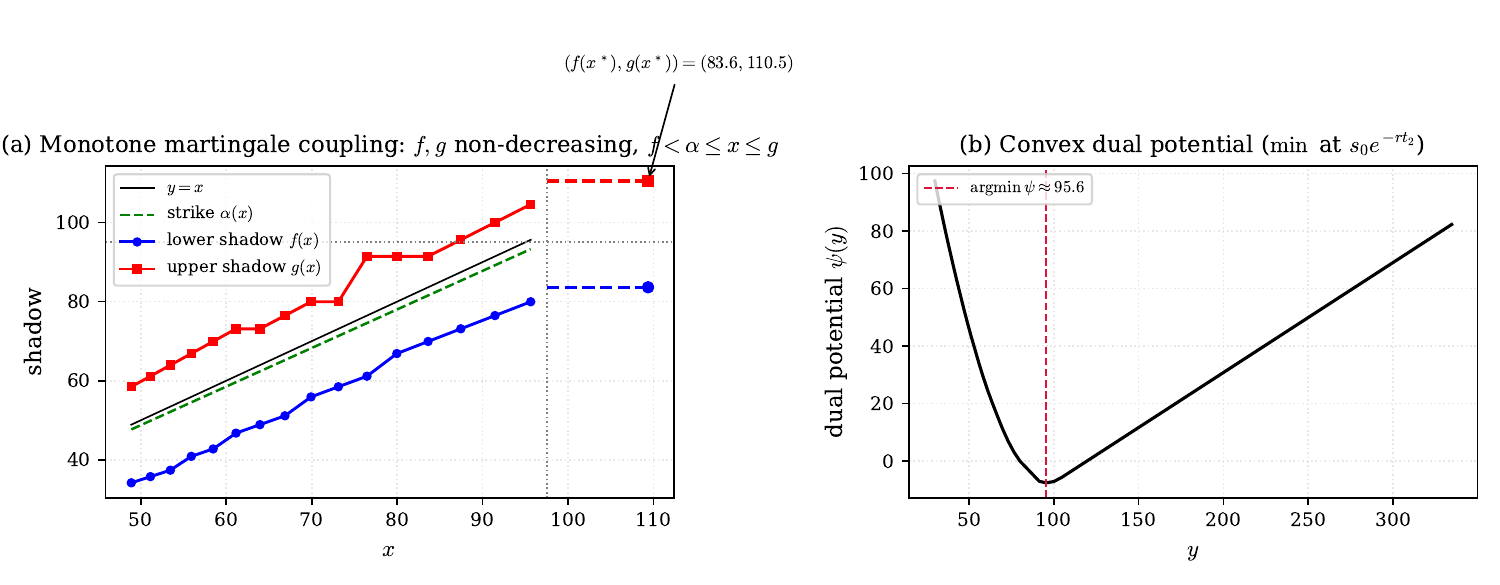}
\caption{(a) The monotone martingale coupling of Proposition~\ref{prop:lc}: both
shadows $f,g$ are non-decreasing with $f<\alpha\le x\le g$, the lower shadow
driven below the strike $\alpha$ (the deep spread). On the plateau the pair
flattens to the boundary value $(f(x^*),g(x^*))=(83.6,110.5)$ of
Corollary~\ref{cor:boundary}, with $g_p$ pinned by the root condition rather than
the degenerate contact condition. (b) The convex dual potential $\psi$ with
$\arg\min\psi\approx s_0e^{-rt_2}$ and $\psi'(f)=-(1+\xi)$, $\psi'(g)=-\xi$.}
\label{fig:shadow}
\end{figure}

\begin{table}[h]
\centering
\begin{tabular}{lcccc}
\toprule
& $n=2$ & $n=3$ & $n=4$ \\
\cmidrule(lr){2-4}
$M$ & $V^{\mathrm{am}}_n\ /\ V^{\mathrm{eu}}_n$ & $V^{\mathrm{am}}_n\ /\ V^{\mathrm{eu}}_n$ & $V^{\mathrm{am}}_n\ /\ V^{\mathrm{eu}}_n$ \\
\midrule
4 & 25.031\ /\ 24.942 & 26.910\ /\ 26.785 & 28.566\ /\ 28.448 \\
6 & 25.836\ /\ 25.719 & 27.843\ /\ 27.700 & 29.589\ /\ 29.430 \\
\bottomrule
\end{tabular}
\caption{Multi-period robust values $V^{\mathrm{am}}_n\ /\ V^{\mathrm{eu}}_n$
(American\ /\ European) from the $n$-marginal regime LP of
Section~\ref{sec:multi}, CRR marginals at $t_i=i\cdot0.5$, $\sigma{=}0.2$,
$r{=}0.05$, $s_0{=}100$, $K{=}120$. The values increase in $n$ (nesting:
the running minimum only ratchets down, so one more period never decreases the
European value, and the American value inherits this) and in $M$ (converging
upward; $V^{\mathrm{am}}_2\to26.90$ at $M{=}40$ and $V^{\mathrm{am}}_3\to29.66$
at $M{=}14$). The continuation coupling is two-point for $n=2$ (the monotone
martingale coupling of Proposition~\ref{prop:lc}) and multi-point---the
mass-diffusion signature of Remark~\ref{rem:n3num}---for $n=3,4$.}
\label{tab:multi}
\end{table}

\section{The multi-marginal extension}\label{sec:multi}

We now indicate how Theorem~\ref{thm:B} extends from two to $n$ prescribed
marginals, at the price of the single structural hypothesis of
\cite{ObSpoida2015}. Let $0<t_1<\dots<t_n$ and let
$\mu_1\le_{\mathrm{cx}}\dots\le_{\mathrm{cx}}\mu_n$ be the prescribed laws of
the discounted prices $X_i=e^{-rt_i}S_{t_i}$; convex order is necessary and
sufficient for a martingale coupling to exist \cite{Strassen65}. Write
$\Pi_m(\mu_1,\dots,\mu_n)$ for the martingale couplings of
$(\mu_1,\dots,\mu_n)$ and
\begin{equation}\label{eq:runmin}
  m_k=\min\bigl(s_0,\,e^{rt_1}X_1,\dots,e^{rt_k}X_k\bigr)
\end{equation}
for the running minimum of the undiscounted price. The robust price of the
American fixed-strike lookback put is
\begin{equation}\label{eq:nprimal}
  P^N=\sup_{\pi\in\Pi_m(\mu_1,\dots,\mu_n)}
       \sup_{\tau\in\mathcal T_n}\E_\pi\bigl[(K-m_\tau)^+\bigr],
\end{equation}
where $\mathcal T_n$ are the $\mathbb F$-stopping times valued in
$\{1,\dots,n\}$. By the continuous-time pricing--hedging duality of
\cite{GuoObloj2025}, $P^N$ equals the minimal superhedging cost over the
two-dimensional state $(X_t,m_t)$.

The running minimum only ratchets \emph{down}, and this is what drives the
structure at every stage. At stage $k$, given the current minimum $m$ and
current price $x$, the holder compares the immediate exercise $(K-m)^+$ with the
continuation value
\begin{equation}\label{eq:ck}
  c_k(x,m)=\sup_{\pi}\E_\pi\Bigl[\bigl(K-\min(m,\,e^{rt_{k+1}}X_{k+1},\dots,e^{rt_n}X_n)\bigr)^+
                 \Bigm| X_k=x\Bigr],
\end{equation}
the supremum over the remaining marginals. The integrand, as a function of the
terminal value, is again a put-type function of the form
$(\alpha(\cdot)-Y)^+$ with $\alpha$ non-decreasing; hence the same
dual envelope-theorem argument as in Proposition~\ref{prop:lc} selects the
\emph{monotone martingale coupling} as the optimal one-step coupling at every
stage, provided the one-step continuation cost inherits the put-type form. This
inheritance is exact at the terminal stage $k=n-1$ (where \eqref{eq:ck} is
precisely the two-period cost of Lemma~\ref{lem:decomp}) and is the content of
the induction step for $k<n-1$; both are made precise in the proof of
Theorem~\ref{thm:multi} below. The multi-marginal object is the iterated
monotone martingale coupling, and its identification with a reflected embedding
rests on the following reflection identity.

\begin{lemma}[Dynamic programming for the multi-marginal value]\label{lem:bellman}
Let $V_k(x,m)$ be the optimal expected discounted (to time $0$) payoff of the
American problem restricted to exercise times $\{t_k,\dots,t_n\}$, started from
$(X_k=x,\ m)$. Then $V_n(x,m)=d_n\,(K-m)^+$ and, for $k<n$,
\begin{equation}\label{eq:bellman}
  V_k(x,m)\;=\;\max\Bigl(d_k\,(K-m)^+,\;
  \sup_{\pi_k}\E_{\pi_k}\bigl[V_{k+1}\bigl(X_{k+1},\min(m,e^{rt_{k+1}}X_{k+1})\bigr)
  \bigm|X_k=x\bigr]\Bigr),
\end{equation}
the inner supremum over one-step martingale couplings of $\delta_x$ into
$\mu_{k+1}$, subject to the aggregate consistency that, mixed over $x\sim\mu_k$,
they reproduce $\mu_{k+1}$. The multi-marginal American value is $V_1(s_0,s_0)$
and the optimal stopping rule exercises at the first stage where the maximum in
\eqref{eq:bellman} is attained by the exercise branch. Hence the multi-period
problem factorises into a chain of one-step martingale optimal transport problems
with cost $g_k^m(\cdot)=V_{k+1}(\cdot,\min(m,e^{rt_{k+1}}\cdot))$ of
Lemma~\ref{lem:inherit}, and the optimal multi-marginal coupling is the
concatenation of the optimal one-step couplings.
\end{lemma}

\begin{proof}
The regime model of \cite{HN2017} attains the model-free American supremum as a
single martingale coupling of the full path together with an adapted exercise
rule. For a fixed rule the value satisfies the Snell-envelope recursion obtained
by conditioning on $X_k$ and updating the running minimum
$m\mapsto\min(m,e^{rt_{k+1}}X_{k+1})$; taking the supremum over rules and
couplings gives \eqref{eq:bellman}, the one-step supremum being exactly the
one-step MOT problem with marginals $\mu_k\le_{\mathrm{cx}}\mu_{k+1}$ and cost
$g_k^m$.
\end{proof}

\begin{lemma}[Reflection: minimum put $\leftrightarrow$ maximum call]\label{lem:reflect}
Under the involution $(S_t)_{t}\mapsto(-S_t)_{t}$ the running minimum
$m_t=\min_{u\le t}S_u$ and the reflected running maximum
$\widetilde M_t:=\max_{u\le t}(-S_u)$ satisfy $\widetilde M_t=-m_t$, so that
\[
  (K-m_t)^+=\bigl(\widetilde M_t+K\bigr)^+,
\]
i.e.\ the fixed-strike lookback put on the running minimum is the fixed-strike
lookback call with strike $-K$ on the reflected running maximum. At the level of
couplings, the involution $(X,Y)\mapsto(-X,-Y)$ sends a martingale coupling with
shadow pair $(f,g)$ to the martingale coupling with shadow pair
$\bigl(-g(-\cdot),\,-f(-\cdot)\bigr)$; in particular the class of monotone
martingale couplings (both shadows non-decreasing) is invariant under this
involution, and the minimum problem is exactly the reflection of the maximum
problem.
\end{lemma}

\begin{assumption}[Ob\l\'oj--Spoida, reflected]\label{ass:star}
The marginals $\mu_1,\dots,\mu_n$ are in convex order and, after centering by
the mean $s_0$, the reflected marginals $\mu_i^-(\cdot)=\mu_i(s_0-\cdot)$
satisfy condition~(ii) of \cite[Ass.\ $\circledast$, \S2.2]{ObSpoida2015}: for
every $2\le i\le n$ and every $y$ in the support of $\mu_i^-$, the map
$\zeta\mapsto c^i(\zeta,y)/(y-\zeta)$ has a unique minimiser on
$[\ell_{\mu_i^-},\,y]$, where $c^i$ is the $i$-th iterated potential of
\cite{ObSpoida2015} (the potential of $\mu_i^-$ penalised by the constraints
carried by $\mu_1^-,\dots,\mu_{i-1}^-$); and $c_i(y)>c_{\iota_i(y;y)}(y)$
whenever the minimiser is strictly below $y$.
\end{assumption}

\begin{lemma}[Put-type inheritance]\label{lem:inherit}
For a fixed current minimum $m$, write the one-step continuation cost of stage
$k<n-1$ as
\begin{equation}\label{eq:inherit}
  g_k^m(y)\;=\;V_{k+1}\!\bigl(y,\;\min(m,e^{rt_{k+1}}y)\bigr),
\end{equation}
where $V_{k+1}$ is the stage-$(k{+}1)$ value function. If for every $m$ the map
$y\mapsto V_{k+1}(y,\min(m,e^{rt_{k+1}}y))$ is convex and non-increasing in
$y$, then $g_k^m$ is a superposition of puts,
\[
  g_k^m(y)\;=\;\int_0^\infty (a-y)^+\;\eta_k^m(\mathrm da)\;+\;\beta_k^m-\gamma_k^m y,
  \qquad \eta_k^m\ge0,\ \gamma_k^m\ge0,
\]
by the convex representation of a non-increasing convex function (its second
derivative is the measure $\eta_k^m$). Moreover the $y$-slope $\partial_y g_k^m$
is non-increasing in $x$: the strike parameter enters only through
$m=\min(\bar m,e^{rt_k}x)$, which is non-decreasing in $x$, and each put
$(a-y)^+$ has slope $-1_{\{y<a\}}$ non-increasing in $a$. Thus the stage-$k$
cost is \emph{convex in $y$} and \emph{submodular}---precisely the two
properties used by the dual envelope argument of Proposition~\ref{prop:lc};
since the put representation is linear in the cost, the argument extends from a
single put to the superposition, and the optimal one-step coupling is the
monotone martingale coupling.
\end{lemma}

\begin{proposition}[One-step coupling: two-point iff convex]\label{prop:onestep}
Fix $k<n$ and a running minimum $m$. The one-step cost
$g_k^m(\cdot)=V_{k+1}(\cdot,\min(m,e^{rt_{k+1}}\cdot))$ of
Lemma~\ref{lem:inherit} is non-increasing in $y$. The optimal one-step coupling
of \eqref{eq:bellman} is
\begin{enumerate}[label=(\roman*)]
\item the monotone two-point martingale coupling of
Proposition~\ref{prop:lc}, whenever $y\mapsto g_k^m(y)$ is convex; and
\item in general, the coupling characterised by the complete duality of
\cite{BNT2015}: for each $x$ the conditional law is supported on the contact set
of $y\mapsto g_k^m(y)-\xi(x)y$ with its convex envelope (the dual potential
$\psi$), where $\xi(x)$ is the martingale multiplier; the support exceeds two
points exactly where $y\mapsto g_k^m(y)$ fails to be convex, and this multi-point
structure is the mass diffusion of Remark~\ref{rem:reflect}.
\end{enumerate}
\end{proposition}

\begin{proof}
Part~(i) is Lemma~\ref{lem:inherit} together with
Proposition~\ref{prop:lc}. For part~(ii), the complete duality of \cite{BNT2015}
gives
$\sup\E[g_k^{m(X)}(Y)]=\inf\{\E_\mu[\varphi]+\E_\nu[\psi]:
\varphi(x)+\psi(y)+\xi(x)(y-x)\ge g_k^{m(x)}(y)\}$; for fixed $x$ the binding
constraint makes $\psi$ the convex envelope of $y\mapsto g_k^{m(x)}(y)$ up to the
affine term $\xi(x)y$, and the optimal conditional law charges only the contact
set. On an affine piece of the envelope the mass is spread across its endpoints,
giving the off-boundary diffusion of \cite[\S4.2]{ObSpoida2015} in the reflected
picture. The argument is local in the stage, so it applies uniformly in $k$, and
the recursion of Lemma~\ref{lem:bellman} closes without any further global
condition.
\end{proof}

\begin{remark}[Scope of the inheritance]\label{rem:inherit}
The convexity-in-$y$ hypothesis is verified at the terminal stage, where
$V_n(y,m')=d_n\bigl(K-\min(m',e^{rt_n}y)\bigr)^+$ is a put in $y$. The
induction closes provided convexity-in-$y$ is preserved by the one-step
maximisation $V_k=\max(\text{exercise},\,c_k)$. This preservation holds under
Assumption~\ref{ass:star} but \emph{fails in general}: the composition
$y\mapsto V_{k+1}(y,\min(m,e^{rt_{k+1}}y))$ need not be convex, and its failure
is exactly the mass diffusion of Remark~\ref{rem:reflect}
(Proposition~\ref{prop:onestep}(ii)), observed numerically
in Remark~\ref{rem:n3num} as a multi-point (not two-point) one-step coupling.
Thus Lemma~\ref{lem:inherit} characterises precisely the ``$\circledast$ holds''
regime in which the iterated coupling is two-point; the unconditional coupling is
the reflected iterated Az\'ema--Yor embedding of step~(iv) in the proof of
Theorem~\ref{thm:multi}. On the interior $\{e^{rt_k}x>\bar m\}$ the
running minimum is the fixed $\bar m$, so the strike is independent of $x$: the
cost is source-independent, the one-step coupling is a free parameter (the
monotone martingale coupling being one extremal choice), and this matches the
two-period plateau degeneracy of Corollary~\ref{cor:boundary}.
\end{remark}

\begin{theorem}[Multi-marginal right-tail structure]\label{thm:multi}
The optimal exercise region at each stage $k<n$ is a right tail
\begin{equation}\label{eq:righttailk}
  \bigl\{\,x\ge x^*_k(m)\,\bigr\},
\end{equation}
where $m$ is the current running minimum and $x^*_k(m)$ is the indifference root
of the stage-$k$ problem \eqref{eq:ck}; the threshold $x^*_k(m)$ is
non-decreasing in $m$, and $x^*_k(s_0)=x^*$ recovers the two-period threshold of
Theorem~\ref{thm:B}. The optimal coupling is the reflected iterated
Az\'ema--Yor embedding of \cite{HOST2016} (step~(iv) of the proof), which
minimises the running minimum in stochastic order; under
Assumption~\ref{ass:star} this is the two-point iterated monotone martingale
coupling (the reflection of the embedding of \cite{ObSpoida2015}), and without it
the coupling carries the mass-diffusion correction of \cite[\S4.2]{ObSpoida2015}.
\end{theorem}

\begin{proof}
We proceed by backward induction, keeping the coupling step and the stopping
step distinct.

\emph{(i) Terminal stage.} At $k=n-1$ the continuation value \eqref{eq:ck} is,
for fixed current minimum $m$, exactly the two-period problem of
Section~\ref{sec:structure} with marginals $\mu_{n-1},\mu_n$: the terminal
payoff $(K-\min(m,e^{rt_n}X_n))^+$ is the two-period continuation payoff
\eqref{eq:payoffs} with the locked minimum $m$ in place of $m_1(x)$. By
Lemma~\ref{lem:reduction} and Proposition~\ref{prop:lc} the optimal one-step
coupling is the monotone martingale coupling, and by Theorem~\ref{thm:B} the
exercise region is a right tail $\{x\ge x^*_{n-1}(m)\}$.

\emph{(ii) Induction: stopping structure.} Assume the stage-$(k{+}1)$ value
$V_{k+1}(y,m')$ is attained by the reflected iterated Az\'ema--Yor embedding of
step~(iv) and has a
right-tail exercise region $\{y\ge x^*_{k+1}(m')\}$, with $x^*_{k+1}(\cdot)$
non-decreasing. For fixed $m$, the immediate exercise value $(K-m)^+$ is
constant in $x$. The continuation value
$c_k(x,m)=\sup_\pi\E_\pi[V_{k+1}(X_{k+1},\min(m,e^{rt_{k+1}}X_{k+1}))\,|\,X_k{=}x]$
is non-increasing in $x$: raising $x$ shifts the law of the next price upwards in
convex order, since the reflected iterated Az\'ema--Yor embedding has
non-decreasing shadow maps and hence stochastically monotone one-step
conditional laws---a property it enjoys in both its two-point and its
mass-diffused form, so it does not rely on Assumption~\ref{ass:star}---while the
effective minimum
$\min(m,e^{rt_{k+1}}X_{k+1})$ only
ratchets down, and $V_{k+1}$ is non-increasing in its minimum argument and---by
the stage-$(k{+}1)$ right-tail structure---non-increasing in the price on the
continuation set. Hence ``continuation minus exercise'' is non-increasing in
$x$, and its unique zero is the indifference root $x^*_k(m)$: the exercise region
$\{x\ge x^*_k(m)\}$ is a right tail. The monotonicity of $x^*_k(m)$ in $m$ is
the same slope comparison as in Theorem~\ref{thm:B}, with the locked intrinsic
$(K-m)^+$ in place of $(K-s_0)^+$.

\emph{(iii) Induction: coupling.} On the continuation set the one-step cost is
$g_k^m(\cdot)$ of \eqref{eq:inherit}. By the stage-$(k{+}1)$ induction hypothesis
$V_{k+1}$ is convex and non-increasing in $y$, so
Lemma~\ref{lem:inherit} applies: $g_k^m$ is a superposition of puts whose strikes
are non-decreasing in $x$ through $m=\min(\bar m,e^{rt_k}x)$, hence the cost is
convex in $y$ and submodular, and the dual envelope argument of
Proposition~\ref{prop:lc}---extended from a single put to the superposition by
linearity---yields the monotone martingale coupling as the optimal one-step
coupling at stage $k$. Applied stage by stage this gives the two-point iterated
monotone martingale coupling as the multi-marginal optimizer \emph{provided} the
convexity-in-$y$ hypothesis of Lemma~\ref{lem:inherit} is preserved at every
stage---equivalently, under Assumption~\ref{ass:star}. This is the only content of
the present step; the convexity-in-$y$ hypothesis \emph{fails} in general, and
its failure is exactly the mass-diffusion correction tracked in
Remark~\ref{rem:n3num}. The general optimality of the coupling is proved in
step~(iv), which does not use the two-point form.

\emph{(iv) Reflection identification (general coupling optimality).} By
Lemma~\ref{lem:reflect} the minimum lookback put is the reflection of the maximum
lookback call, so the optimal coupling for the former is the reflection of the
optimal coupling for the latter. By the sharp maximum--maximum bound of
\cite{HOST2016}, the iterated Az\'ema--Yor embedding maximises the running
maximum in stochastic order among all martingales with marginals
$(\mu_1^-,\dots,\mu_n^-)$; reflecting, this embedding minimises the running
minimum in stochastic order. This global statement is consistent with the local
recursion of Lemma~\ref{lem:bellman}: at each stage the one-step optimum is
characterised by Proposition~\ref{prop:onestep}, which is the two-point coupling
under Assumption~\ref{ass:star} and carries the mass diffusion otherwise. Since
$(K-m_\tau)^+$ is a non-increasing function of
the running minimum, the reflected iterated Az\'ema--Yor embedding is optimal for
the American value \eqref{eq:nprimal}---this holds unconditionally, with no
reliance on Assumption~\ref{ass:star}. Under Assumption~\ref{ass:star} the
iterated Az\'ema--Yor embedding of \cite{ObSpoida2015} is the two-point iterated
monotone coupling of step~(iii) (by \cite[Thm.~2.3]{ObSpoida2015} it is explicit
and realises the law of the maximum at every stopping time); without it, the
embedding carries the mass-diffusion correction of \cite[\S4.2]{ObSpoida2015},
which is multi-point and is not of the two-point monotone form.
\end{proof}

\begin{remark}[Reflection and the maximum-maximum]\label{rem:reflect}
The minimum lookback put is the reflection of the maximum lookback call:
$(K-\min_i S_{t_i})^+$ is a function of the running maximum of the reflected
process $-S$. The two-period monotone martingale coupling is the reflection of
the two-period Az\'ema--Yor (right) barrier, and the multi-marginal statement
above is the reflection of the ``maximum-maximum'' bound of \cite{HOST2016}; this
is why Assumption~\ref{ass:star} is imposed on the reflected marginals
$\mu_i^-$. As in \cite[\S4]{ObSpoida2015} the assumption is not necessary: when
Assumption~$\circledast$(ii) fails, the iterated Az\'ema--Yor embedding must be
amended by a \emph{mass diffusion} between successive stopping times---the
second stopping time is replaced by one which diffuses the mass stopped at a
boundary point to two off-boundary points \emph{without changing the running
maximum}---and the optimisation determining the next boundary gains an explicit
``error'' book-keeping term. This is carried out in \cite[\S4.2]{ObSpoida2015}
for $n=3$, where the modified boundary $\tilde\xi_3$ and the error term are made
explicit, and it is not extended to $n>3$. In reflected (minimum) form this is
precisely the correction to the iterated monotone martingale coupling when the
reflected marginals violate $\circledast$(ii); the quantitative analysis of
these correction terms for general $n$ is a self-contained constructive
programme, best treated in a separate follow-up paper. By
Remark~\ref{rem:telescope}
the correction is local, so the $n>3$ construction is the iteration of the
$n=3$ operator of \cite[\S4.2]{ObSpoida2015} rather than a new object.
\end{remark}

\begin{remark}[Numerical confirmation of mass diffusion ($n=3$)]\label{rem:n3num}
The three-marginal regime LP of Section~\ref{sec:multi}---the direct
generalisation of the two-period LP, with one regime per exercise stage and each
regime separately martingale---was solved for CRR marginals
$(\mu_1,\mu_2,\mu_3)$ at $t=(0.5,1.0,1.5)$ with $\sigma=0.2$, $r=0.05$,
$s_0=100$, $K=120$. The value is $V_3\approx 29.66$ (grid $3M{+}1$ nodes,
$M=14$, still increasing in $M$; see Table~\ref{tab:multi}), against $V_2\approx 26.87$ for the two-period
problem and $\approx 28.84$ for the three-period European value, and the
first-stage exercise region is again a right tail with threshold
$x^*_1\approx 109$. Two structural facts emerge. \emph{(i) The mass exercising at
$t_2$ uses the two-point monotone martingale coupling}, exactly as in the
two-period problem. \emph{(ii) The mass continuing to $t_3$ does not:} its
conditional law of $X_2$ given $X_1$ has support of size $10$--$12$ grid points,
not $2$, and this is stable under grid refinement ($M=10,12,14$) and under a
$10^{-4}$ relative perturbation of the payoff, so it is not an LP degeneracy.
The same multi-point structure is already present in the \emph{European} $n=3$
coupling (hold to $t_3$, no stopping), whose conditional law of $X_2$ given
$X_1$ has support of size $2$--$8$; hence the diffusion is generic, not an
artefact of the American stopping freedom, and confirms that the reflected
marginals here violate Assumption~\ref{ass:star}. This is precisely the
mass-diffusion signature of \cite[\S4.2]{ObSpoida2015}: the
iterated monotone coupling is amended by a diffusion of the mass stopped at a
boundary, and the naive two-point iterated monotone coupling of
Theorem~\ref{thm:multi} step~(iii) is strictly suboptimal for $n=3$. The
multi-marginal right-tail \emph{exercise} structure survives; it is the
\emph{coupling} step, not the stopping step, that carries the correction.
\end{remark}

\begin{figure}[htbp]
\centering
\includegraphics[width=\textwidth]{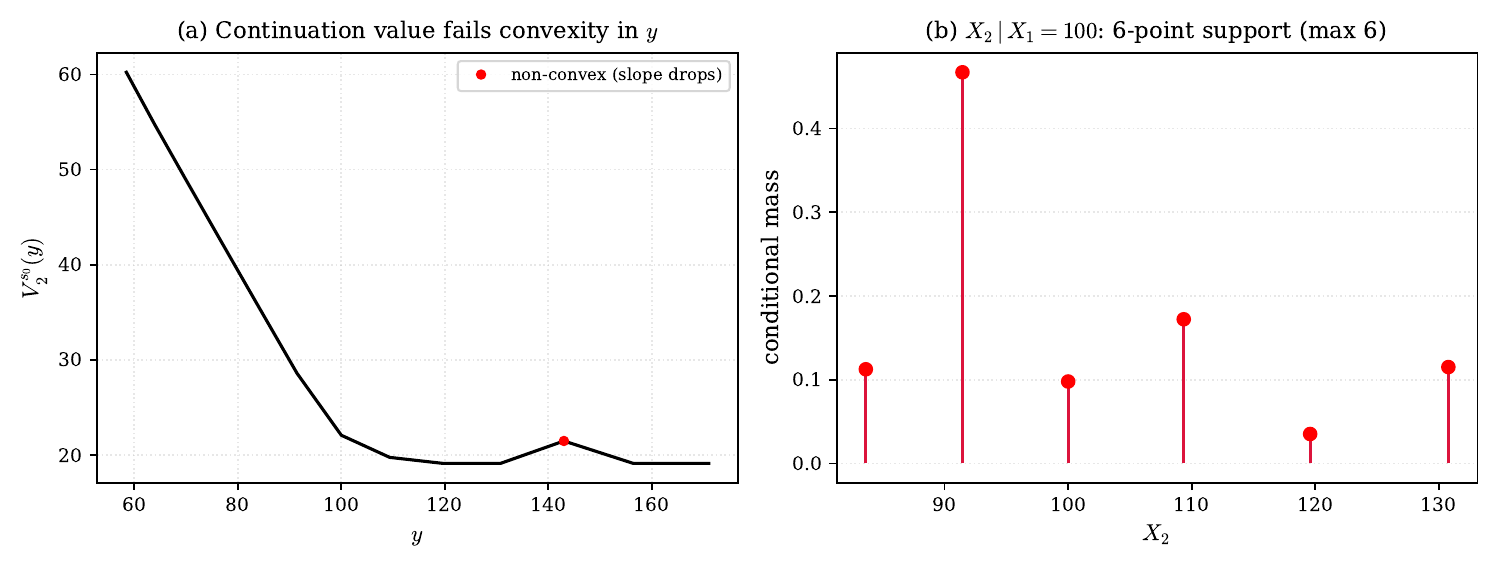}
\caption{Mass diffusion in the three-period problem (Remark~\ref{rem:n3num}).
(a) The continuation value $y\mapsto V_2^{s_0}(y)$ fails convexity (red nodes
where the slope drops). (b) The optimal first-stage conditional law
$X_2\,|\,X_1$ has six support points rather than the two-point monotone
martingale coupling---the signature of the mass-diffusion correction of
\cite[\S4.2]{ObSpoida2015}.}
\label{fig:massdiff}
\end{figure}

\begin{remark}[Direct verification of the one-step mechanism]\label{rem:onestep-num}
The bifurcation of Proposition~\ref{prop:onestep} is directly visible in the LP
data. For the \emph{two-period} problem the first-stage cost is the convex
put-superposition $y\mapsto d_2(K-\min(s_0,e^{rt_2}y))^+$; its finite-difference
slopes on the (geometric, hence non-uniform) CRR grid are non-decreasing---the
correct discrete convexity test there---and the optimal continuation coupling is
the two-point monotone martingale coupling of Proposition~\ref{prop:lc}. For the
\emph{three-period} problem the first-stage cost is the continuation value
$y\mapsto V_2(y,\min(s_0,e^{rt_2}y))$ of Lemma~\ref{lem:bellman}, which
\emph{fails} convexity: its slopes are not monotone, with several strictly
negative second divided differences. The optimal first-stage conditional law
$X_2|X_1$ accordingly has $6$--$8$ support points rather than two ($M=6,8$), and
this multi-point support is stable under random and quadratic perturbations of
the payoff, so it is the genuine mass diffusion of
Proposition~\ref{prop:onestep}(ii)---the contact-set coupling---not an LP
degeneracy. The precise location of the convexity failure is sensitive to the
OTM degeneracy of the two-period sub-problem (the flat dual potential of
Proposition~\ref{prop:lc}), but its \emph{existence}, and the resulting
multi-point support, are robust.
\end{remark}

\begin{lemma}[Consistency of the iterated embedding]\label{lem:consistent}
The iterated Az\'ema--Yor embedding of \cite{HOST2016}---amended, when
Assumption~\ref{ass:star}(ii) fails, by the mass-diffusion correction of
\cite[\S4.2]{ObSpoida2015}---is \emph{consistent under projection}: for every
$k<n$, its marginal on $(X_1,\dots,X_k)$ is the $k$-stage iterated embedding.
Equivalently the optimal $k$-stage coupling does not depend on the marginals
$\mu_{k+1},\dots,\mu_n$, and the correction applied at stage $k$ is a function of
the local marginal triple $(\mu_{k-1},\mu_k,\mu_{k+1})$ alone.
\end{lemma}

\begin{proof}
The embedding is \emph{iterated} by construction: $\mu_1$ is embedded into
$\mu_2$, the resulting law into $\mu_3$, and so on, the transition at stage $k$
depending only on the state $(X_k,m_k)$ and the next marginal $\mu_{k+1}$. When
Assumption~\ref{ass:star}(ii) fails, the mass-diffusion amendment of
\cite[\S4.2]{ObSpoida2015} alters only the local transition between the $k$-th
and $(k{+}1)$-th stopping times, leaving earlier stages untouched. Projecting
the $n$-stage construction onto the first $k$ coordinates therefore returns the
$k$-stage construction.
\end{proof}

\begin{remark}[Numerical confirmation of consistency ($n=4$)]\label{rem:telescope}
Lemma~\ref{lem:consistent} reduces the explicit $n>3$ mass-diffusion question to
the \emph{iteration} of the explicit $n=3$ operator of \cite[\S4.2]{ObSpoida2015}:
future marginals do not perturb earlier stages, so no new global phenomenon
appears. We confirm the consistency at machine precision for $n=4$. The
four-marginal regime LP---the direct extension of the $n=3$ LP of
Remark~\ref{rem:n3num} to four stages $t=(0.5,1.0,1.5,2.0)$, $\sigma=0.2$,
$r=0.05$, $s_0=100$, $K=120$---has American value $V_4\approx29.59$ and European
value $\approx29.43$ (grid $4M{+}1$ nodes, $M=6$; see Table~\ref{tab:multi}). The optimal \emph{European}
four-marginal coupling, projected onto $(X_1,X_2,X_3)$, attains the
three-marginal European optimum under the three-period payoff to within
$10^{-13}$ for $K=100,110,120,140$ (gaps $5\times10^{-14}$, $5\times10^{-14}$,
$4\times10^{-14}$, $6\times10^{-14}$), i.e.\ the projection is an optimal $n=3$
coupling. Moreover the conditional law of $X_4$ given $X_3$ in the $n=4$ coupling
has the same multi-point support (five points) as the $n=3$ case of
Remark~\ref{rem:n3num}, so the mass-diffusion signature reappears one stage
later, exactly as the local-iteration picture predicts.
\end{remark}

\section*{Conclusion}
We have isolated the American fixed-strike lookback put as the non-trivial
American lookback problem, proved the pricing--hedging duality, and established
the right-tail exercise structure in the two-period model, backed by LP
experiments. The pointwise continuation coupling is the monotone martingale coupling
of Proposition~\ref{prop:lc}, whose lower/upper shadows $(f,g)$ are both
non-decreasing with the lower shadow driven below the strike, characterised by
the convex dual potential of \eqref{eq:msdual}; the exercise threshold is fixed
by the indifference condition
\eqref{eq:indiff}. The multi-marginal extension (Section~\ref{sec:multi})
carries this over to $n$ marginals: under the reflected
Ob\l\'oj--Spoida Assumption~\ref{ass:star} the optimal coupling is the iterated
monotone martingale coupling and the exercise region is a right tail at every stage,
Theorem~\ref{thm:multi}. The $n=3$ LP of
Remark~\ref{rem:n3num} confirms the right-tail exercise structure and shows that,
beyond two periods, the optimal coupling carries a genuine mass-diffusion
correction of \cite[\S4.2]{ObSpoida2015}---the stopping (exercise) structure is
clean; the coupling step is where the multi-marginal correction lives. By
Remark~\ref{rem:telescope} this correction is local (consistent under
projection), so the explicit $n>3$ construction is the iteration of the explicit
$n=3$ operator rather than a new object. The explicit book-keeping of the $n>3$
correction terms is a constructive programme of independent interest, which we
leave to a follow-up paper.

\end{document}